\documentclass[10pt,a4paper,twocolumn,fleqn]{article}

\usepackage[authoryear]{natbib}
\usepackage{amsmath}
\usepackage{booktabs}
\usepackage{graphicx}
\usepackage{array}
\usepackage[version=4]{mhchem}
\usepackage{siunitx}
\usepackage{microtype}
\usepackage[percent]{overpic}
\usepackage[hidelinks]{hyperref}
\usepackage[margin=1.8cm,columnsep=0.7cm]{geometry}

\DeclareSIUnit{\angstromunit}{\text{\AA}}
\DeclareSIUnit{\atomunit}{atom}
\newcommand{\kappalatt}{\ensuremath{\kappa_{\mathrm{latt}}}}
\title{Lattice Thermal Conductivity of \ce{MgO} and the \ce{MgSiO3}
Polymorphs Bridgmanite and Post-Perovskite at Lower-Mantle Conditions
Using Fine-Tuned MACE Potentials}

\author{Amrendra Kumar Tiwari,\quad
Sandro Jahn\thanks{Corresponding author:
\href{mailto:s.jahn@lmu.de}{s.jahn@lmu.de}}\\[0.5ex]
\small Department of Earth and Environmental Sciences, LMU Munich\\
\small Faculty of Geosciences, Theresienstra\ss e 41, 80333 Munich, Germany}
\date{}

\begin{document}
\maketitle

\begin{abstract}
Heat flow across the core--mantle boundary (CMB) depends on the lattice
thermal conductivity (\kappalatt) of lowermost-mantle minerals, yet calculated
CMB values for \ce{MgO} differ severalfold, and high-temperature estimates of
the conductivity increase across the bridgmanite--post-perovskite transition
range from about 20 to 50\%. Direct solutions of the linearized phonon
Boltzmann transport equation (LBTE) require thousands of density functional
theory (DFT) force evaluations per thermodynamic state. Here we replace these
evaluations with phase-specific, fine-tuned MACE potentials for \ce{MgO},
\ce{MgSiO3} bridgmanite (Pv), and \ce{MgSiO3} post-perovskite (PPv). The
potentials reproduce DFT energies, forces, stresses, pressure--volume
relations, and compressed-state phonons for all three phases. For \ce{MgO} at
\SI{0}{\giga\pascal}, replacing MACE forces with DFT forces in otherwise
identical transport calculations changes \kappalatt{} by 0.2\% at
\SI{300}{\kelvin} and 1.9\% at \SI{2000}{\kelvin}. PPv is 44--55\% more
conductive than Pv at the same pressure and temperature and strongly
anisotropic, with $\kappa_{[100]}/\kappa_{[010]}=1.7$--2.1. The
quasi-harmonic phase boundary has a Clapeyron slope of
\SI{7.4 \pm 0.1}{\mega\pascal\per\kelvin} and intersects a model geotherm at
\SI{116.2}{\giga\pascal} and \SI{2660}{\kelvin}, where the transition raises
the silicate conductivity by 47\%, to
\SI{8.6}{\watt\per\metre\per\kelvin}. Heating in the basal thermal boundary
layer lowers the \ce{MgO} and PPv conductivities to 37.3 and
\SI{6.7}{\watt\per\metre\per\kelvin} at the CMB, where Hashin--Shtrikman
bounds of \SIrange{9.5}{11.4}{\watt\per\metre\per\kelvin} for an Fe-free
aggregate of 80~vol\% silicate and 20~vol\% \ce{MgO} encompass the
\SI{10}{\watt\per\metre\per\kelvin} used in earlier heat-flow estimates. The
density--temperature parameterizations provide an intrinsic end-member
reference for lower-mantle thermal models.
\end{abstract}

\noindent\textbf{Keywords:} lattice thermal conductivity; core--mantle
boundary heat flow; post-perovskite phase
transition; periclase; machine-learned interatomic potentials; phonon
Boltzmann transport equation

\section{Introduction}

Heat conducted across the core--mantle boundary (CMB) couples the thermal
evolution of Earth's core and mantle. It governs how fast the core cools and
the inner core grows, how much energy is available to the geodynamo, and how
strongly the lowermost mantle is heated from below. Converting the temperature
contrast across the basal thermal boundary layer into a heat flux requires the
thermal conductivity of lowermost-mantle minerals at pressures near
\SI{135}{\giga\pascal} and temperatures up to
about \SI{4000}{\kelvin}. Measurements under such conditions are still rare. Thermoreflectance
experiments on \ce{MgO} have reached about \SI{140}{\giga\pascal} at
\SI{1950}{\kelvin}, and about \SI{3170}{\kelvin} at lower pressures
\citep{Hasegawa2023}. Recent work on bridgmanite and
Fe-bearing post-perovskite has also extended to high pressures and
temperatures, but not yet across the full range expected near the CMB
\citep{Edmund2024,Okuda2020}. Atomistic calculations therefore remain central
both to interpreting these experiments and to extending them to deep-mantle
conditions. Here we consider only the lattice contribution to the thermal conductivity
of the Fe-free end members \ce{MgO}, \ce{MgSiO3} bridgmanite, and
\ce{MgSiO3} post-perovskite. Radiative and electronic transport,
chemical substitution, point defects, and grain boundaries must be added
separately when the results are applied to natural mantle assemblages.

\ce{MgO} is the Fe-free end member of \ce{(Mg,Fe)O} ferropericlase, the second most abundant
phase in a pyrolitic lower mantle, and \ce{MgSiO3} bridgmanite is the most
abundant. Because of its simple rocksalt structure, \ce{MgO} has become a
standard benchmark for high-pressure thermal transport. Bridgmanite, with an
orthorhombic \textit{Pbnm} structure, has a much richer and
direction-dependent phonon spectrum. Near the base of the mantle, it
transforms to \ce{MgSiO3} post-perovskite, a \ce{CaIrO3}-type phase with
\textit{Cmcm} symmetry
\citep{Murakami2004,OganovOno2004,TsuchiyaT2004}. In post-perovskite,
edge- and corner-sharing \ce{SiO6} octahedra form sheets separated by
Mg-bearing interlayers. The phase is only about 1--1.5\% denser than
bridgmanite, but its layered structure makes it strongly anisotropic, with
the $b$ axis, normal to the sheets, by far the most compressible direction
\citep{Guignot2007,Iitaka2004,Komabayashi2008,Murakami2004}.

Where exactly the transition occurs is still uncertain. In situ experiments
at temperatures up to \SI{4380}{\kelvin} yielded a Clapeyron slope of
\SI{13.3 \pm 1.0}{\mega\pascal\per\kelvin} and a transition temperature of
\SI{3520 \pm 70}{\kelvin} at CMB pressure \citep{Tateno2009}.
First-principles studies generally predict lower slopes of approximately
\SIrange{7.5}{10}{\mega\pascal\per\kelvin}
\citep{OganovOno2004,TsuchiyaT2004,TsuchiyaJ2005,ZhangWentzcovitch2022},
and phonon-quasiparticle calculations show that intrinsic anharmonicity
curves the boundary at high temperature \citep{ZhangWentzcovitch2022}.
Because the phase boundary lies within the pressure--temperature range of the
D$^{\prime\prime}$ layer, a geotherm that steepens in the basal thermal
boundary layer may cross it twice and produce paired seismic discontinuities
of opposite polarity \citep{Hernlund2005}. The depths of these discontinuities
have been used to constrain the temperature gradient above the CMB. Converting
that gradient into a heat flow, however, has relied on an assumed thermal
conductivity of \SI{10}{\watt\per\metre\per\kelvin} for the
D$^{\prime\prime}$ layer \citep{Hernlund2005,Tateno2009}. The conductivity
contrast between bridgmanite and post-perovskite enters these estimates
directly.

For \ce{MgO}, experiment and theory agree that the lattice conductivity rises
under compression and falls with temperature, but not on its magnitude at
CMB conditions. Theoretical estimates there range from roughly
\SIrange{20}{80}{\watt\per\metre\per\kelvin}, depending on the interatomic
description and on how anharmonicity is treated
\citep{Chabane2025,deKoker2010,DekuraTsuchiya2017,Haigis2012,
Stackhouse2010,TangDong2010}. Most experiments have isolated either the
pressure or the temperature dependence. The measurements of
\citet{Hasegawa2023} are an exception and constrain the conductivity directly
at simultaneously high pressure and temperature. The way anharmonicity is
handled can, on its own, shift the result considerably. \citet{Chabane2025} found that calculations including three-phonon
scattering but not the accompanying higher-order renormalization of phonon
frequencies underestimate the conductivity of \ce{MgO} by about 20--45\%
along the lower-mantle geotherm. Deep-potential Green--Kubo simulations of
\ce{MgO} up to super-Earth conditions likewise indicate that four-phonon
scattering is significant and that its contribution relative to
three-phonon scattering varies with density \citep{Qiu2025}. Isotopic
disorder, oxygen vacancies, and Fe substitution lower the conductivity
further \citep{Chabane2025}, so ideal \ce{MgO} should be read as a
high-conductivity end member, not as a model for natural ferropericlase.

Bridgmanite is less well constrained. Anharmonic lattice-dynamics calculations
differ widely in absolute values. At \SI{100}{\giga\pascal}
and \SI{300}{\kelvin}, \citet{Dekura2013} obtained
\SI{28.1}{\watt\per\metre\per\kelvin}. \citet{Ghaderi2017}, who solved the
linearized Boltzmann equation directly and tested convergence with respect to
the range of the anharmonic interactions, obtained
\SI{59.2}{\watt\per\metre\per\kelvin}. The temperature dependence is equally
method-sensitive. Third-order lattice-dynamics calculations approach a
$T^{-1}$ law \citep{Tang2014}, multianvil measurements gave $T^{-0.43}$
\citep{Manthilake2011}, and recent optical- and X-ray-heating experiments
point to exponents of only about 0.2--0.5 \citep{Edmund2024}. Molecular
dynamics driven by machine-learned potentials reaches far larger systems and
longer trajectories than ab initio molecular dynamics and includes
anharmonicity beyond third order
\citep{PengDeng2024,Qiu2025,Wang2023,Yang2022}. It also brings its own concerns,
namely finite-size effects, statistical convergence, and the classical
treatment of the nuclei.

The contrast between bridgmanite and post-perovskite is even less certain.
\citet{Ohta2012} measured Fe-free polycrystalline samples at
\SI{300}{\kelvin} and found post-perovskite at
\SIrange{135}{141}{\giga\pascal} to be about 72\% more diffusive than
bridgmanite. Measurements on the structural analogue \ce{CaIrO3} at
300--600~\si{\celsius} show a similar increase, by a factor of about 1.8
\citep{Hunt2012}. Absolute values
from diamond-anvil-cell samples need care at room temperature, however, where
grain-boundary scattering is strongest. \citet{Okuda2020} estimated that it
lowers the \SI{300}{\kelvin} conductivity of fine-grained post-perovskite by
about 61\%, but by much less for mantle grain sizes. The same study measured
Fe-bearing post-perovskite up to \SI{1558}{\kelvin}. Comparable
high-temperature data for the Fe-free end member are still lacking.

Calculations at high temperature generally find post-perovskite to be the
better conductor, but they disagree on how much better. Nonequilibrium
molecular dynamics with classical interatomic potentials gave an increase of
about 50\%, with the conductivity along the $a$ axis about 40\% higher than
along the $c$ axis \citep{Ammann2014}. A direct solution of the phonon
Boltzmann equation likewise gave about 50\% and identified the $a$ and $b$
axes as the most and least conductive directions, respectively
\citep{DekuraTsuchiya2019}. Smaller contrasts were obtained from
phonon-quasiparticle calculations, about 25\% \citep{ZhangWentzcovitch2021},
and from deep-potential molecular dynamics, 18.2\% along a model geotherm
\citep{Yang2022}; for a pyrolitic assemblage, \citet{Wang2023} obtained about
22\%. These studies differ in interatomic interactions, composition,
temperature, and transport method, so they cannot be compared one to one.
Taken together, they leave the magnitude of the conductivity increase across
the transition, and the degree of post-perovskite anisotropy, among the least
constrained inputs to models of heat transport through the D$^{\prime\prime}$
layer.

Anharmonic lattice dynamics combined with the linearized phonon Boltzmann
transport equation (LBTE) offers a mode-resolved way to address this problem.
Harmonic (second-order) force constants, the second derivatives of the
potential energy with respect to atomic displacements, determine the phonon
frequencies, eigenvectors, and group velocities. Cubic (third-order) force
constants, the third derivatives, set the strength of three-phonon
interactions and, together with the harmonic phonons, determine the
three-phonon scattering rates. A direct solution of the LBTE retains the
coupling between phonon modes that is neglected in the single-mode
relaxation-time approximation \citep{Chaput2013,Togo2015}. Different
transport packages give nearly identical conductivities when supplied with
the same force constants. When complete workflows were run independently,
the results agreed to within about \(\pm15\%\), with most of the remaining
spread arising from differences in the cubic force constants and the
resulting phonon lifetimes \citep{McGaughey2025}. Accurate, well-converged
cubic force constants therefore matter at least as much as the choice of
solver. They are also the main computational burden, because low-symmetry
silicates can require thousands of supercell force calculations at every
thermodynamic state.

Machine-learned interatomic potentials can remove most of this cost while
leaving the transport theory unchanged. MACE describes atomic environments
through equivariant message passing with high-body-order features and
evaluates energies and forces at a small fraction of the cost of density
functional theory (DFT)
\citep{Batatia2022}. Foundation models pretrained on broad chemical datasets
can then be fine-tuned for a given material and pressure--temperature range
\citep{Batatia2025}. At lower-mantle pressures such adaptation is necessary.
Universal potentials used without fine-tuning can systematically soften
phonon frequencies \citep{Loew2025}, and their errors grow under compression
when high-pressure structures are scarce in the training data
\citep{Loew2026}. Fine-tuning has reduced force errors by factors of about
5--15 across diverse systems \citep{Hanseroth2026}. Low force errors alone
are not proof of suitability, however. \citet{Pota2026} showed that a
reasonable macroscopic conductivity can result from errors in harmonic and
anharmonic properties that partly cancel. A model intended for
thermal-conductivity calculations has to be tested against the quantities
that actually control heat transport.

Previous lower-mantle applications of machine-learned potentials have mainly
combined them with molecular dynamics \citep{PengDeng2024,Qiu2025,Wang2023,Yang2022},
whereas phonon and thermal-transport calculations with MACE have so far been
benchmarked largely at ambient pressure using general-purpose foundation
models \citep{Pota2026}. Two linked questions therefore motivate this study.
First, can system-specific, fine-tuned MACE potentials replace direct DFT
force evaluations for the large finite-displacement sets required by
direct-LBTE calculations at lower-mantle conditions without compromising the
resulting transport properties? Second, when \ce{MgO}, bridgmanite, and
post-perovskite are treated within a common transport framework, what do
their relative conductivities and the directional anisotropy of the
silicates, together with the conductivity change across the
bridgmanite--post-perovskite transition, imply for heat transport through
the lowermost mantle? We address these questions with phase-specific
potentials fine-tuned from the same MACE foundation model. To our knowledge,
this is the first application of system-specific, fine-tuned MACE potentials
to direct-LBTE calculations of major lower-mantle minerals across the
relevant pressure--temperature range. We validate the potentials against
independent DFT energies, forces, stresses, quasi-harmonic pressure--volume
relations, and compressed-state phonons, and, for \ce{MgO}, directly at the
level of the transport calculation. We then use DFT-QHA Gibbs free energies
to locate the bridgmanite--post-perovskite boundary, quantify the phase and
directional conductivity contrasts, derive density--temperature
parameterizations, and evaluate phase and aggregate conductivities along a
model lower-mantle geotherm.

\section{Computational details}
\label{sec:methods}

\subsection{DFT reference calculations and training databases}
\label{sec:dft}

DFT calculations were performed with VASP using the
projector-augmented-wave (PAW) method and the Perdew--Burke--Ernzerhof
(PBE) generalized gradient approximation
\citep{KresseFurthmuller1996,KresseJoubert1999,Perdew1996}. The reference
databases were generated using plane-wave cutoffs of
\SI{600}{\electronvolt} for \ce{MgO} and \SI{650}{\electronvolt} for both
\ce{MgSiO3} polymorphs, bridgmanite (Pv) and post-perovskite (PPv).
Brillouin-zone sampling was converged for each simulation cell. The PAW
datasets, $k$-point meshes, electronic convergence criteria, and ab initio
molecular dynamics (AIMD) settings are listed in Table~S1.

A reference database was constructed for each phase following the same
strategy. AIMD trajectories sampled the pressure--temperature range of
interest. Because consecutive AIMD frames are strongly correlated, snapshots
were selected by farthest-point sampling in the space of smooth overlap of
atomic positions (SOAP) descriptors
\citep{Bartok2013SOAP,Imbalzano2018}. This procedure retained structurally
diverse configurations while discarding redundant frames. Additional static
configurations generated by homogeneous strain, random atomic displacements,
systematic changes in volume, and finite atomic displacements sampled the
local distortions relevant to the harmonic and cubic force constants. Every
selected configuration was labelled with its DFT energy, atomic forces, and
stress tensor.

The databases were divided into training, validation, and test subsets by
groups rather than by individual configurations. All selected snapshots from
a given AIMD trajectory and all members of a strain, volume, or displacement
series were assigned to the same subset, preventing closely related
configurations from being split across subsets. Database sizes, partitions,
and thermodynamic coverage are summarized in
Table~\ref{tab:reference_databases}; further details of database construction
are provided in Text~S1 and Table~S2.

\begin{table*}[t]
\centering
\small
\caption{DFT reference databases used to fine-tune the phase-specific MACE
potentials.}
\label{tab:reference_databases}
\begin{tabular}{lcccc}
\toprule
Phase & Configurations & Train/validation/test & $P$ (GPa) & $T$ (K) \\
\midrule
\ce{MgO}         & 3528 & 2822/353/353 & 0--135   & 300--3500 \\
\ce{MgSiO3} Pv  & 4612 & 3758/470/384 & 0--135   & 300--3500 \\
\ce{MgSiO3} PPv & 3575 & 2879/348/348 & 100--145 & 300--4000 \\
\bottomrule
\end{tabular}
\end{table*}

\subsection{MACE fine-tuning and validation}
\label{sec:mace}

Phase-specific potentials for \ce{MgO}, Pv, and PPv were obtained by
fine-tuning the same MACE-MPA-0 medium foundation model using MACE~0.3.16
under a common protocol \citep{Batatia2022,Batatia2025}. The foundation model
was pretrained on the combined MPtrj and sAlex datasets
\citep{Deng2023,BarrosLuque2026}. Each potential retained the pretrained
architecture and was fine-tuned on its own DFT database without replay of the
pretraining data. Energies, forces, and stresses were fitted jointly in double
precision, with relative energy--force--stress loss weights of
$1{:}10{:}100$. For all three potentials, the validation errors reached a
plateau within the 1000-epoch training limit. The remaining optimization
settings are listed in Text~S1 and Table~S1. The test subsets were used
neither for training nor for model selection.

The potentials were validated at three levels. First, MACE energies, forces,
and stresses were compared with DFT for the held-out test configurations.
Second, properties not fitted directly were compared with independent DFT
calculations: quasi-harmonic pressure--volume relations at
\SI{300}{\kelvin} and \SI{2000}{\kelvin} and compressed-state phonon
dispersions for all three phases. Third, for \ce{MgO} at
\SI{0}{\giga\pascal}, \kappalatt\ was recalculated at
\SI{300}{\kelvin} and \SI{2000}{\kelvin} using DFT forces on the same
finite-displacement structures, with all other transport settings unchanged.
This comparison provides a direct transport-level test of how the residual
force errors propagate through the force constants into \kappalatt.

\subsection{Thermal volumes and interatomic force constants}
\label{sec:fc}

Thermal expansion was included using pressure- and temperature-dependent
equilibrium volumes obtained from DFT within the quasi-harmonic approximation
(QHA). At each pressure and temperature, the equilibrium volume of phase $i$
was obtained by minimizing its Gibbs free energy,
\begin{equation}
G_i(P,T)
=
\min_V
\left[
E_{\mathrm{static},i}(V)
+
F_{\mathrm{vib},i}(V,T)
+
PV
\right],
\label{eq:gibbs_qha}
\end{equation}
where $E_{\mathrm{static},i}(V)$ is the static DFT energy and
$F_{\mathrm{vib},i}(V,T)$ is the harmonic vibrational free energy, including
the zero-point contribution. Electronic thermal excitations were neglected.
The calculated pressure--volume--temperature relations, thermal
equation-of-state parameters, and interpolation procedures are described in
Text~S2 and Table~S4. The resulting equilibrium structures were used
consistently in the force-constant and transport calculations.

Second- and third-order interatomic force constants were constructed by
finite displacement using Phono3py~4.4.0
\citep{Togo2015,Togo2023}. Forces for the displaced supercells were evaluated
with the corresponding fine-tuned MACE potential using a displacement
amplitude of \SI{0.03}{\angstromunit}; no pair-distance cutoff was imposed on
the third-order interactions. The production supercells were a
$5\times5\times5$ repetition of the two-atom \ce{MgO} primitive cell and
$2\times2\times2$ and $4\times2\times2$ repetitions of the 20-atom
conventional cells of Pv and PPv, respectively. They contained 250, 160, and
320 atoms and required 842, 14\,675, and 15\,440 symmetry-reduced third-order
displacement configurations, respectively. Evaluating forces for these
configurations at every thermodynamic state is the principal computational
bottleneck of a direct-DFT workflow and is the step in which MACE replaced
direct DFT force evaluations.

Non-analytical term corrections were included for all three phases using Born
effective-charge tensors and high-frequency dielectric tensors obtained from
DFT linear-response calculations for the two-atom \ce{MgO} cell and the
20-atom Pv and PPv cells at the corresponding equilibrium structures. The
phase-specific $k$-point meshes are listed in Table~S1.

\subsection{Lattice thermal conductivity}
\label{sec:kappa}

The same transport treatment was applied to all three phases. The lattice
thermal-conductivity tensor, $\kappa_{\mathrm{latt},\alpha\beta}$, was
obtained by direct solution of the LBTE following \citet{Chaput2013}, as
implemented in Phono3py \citep{Togo2015,Togo2023}:
\begin{equation}
\kappa_{\mathrm{latt},\alpha\beta}
=
\frac{1}{N_{\mathbf q}V}
\sum_{\lambda}
C_{\lambda}
v_{\lambda,\alpha}
F_{\lambda,\beta},
\label{eq:kappa_tensor}
\end{equation}
where $\alpha,\beta\in\{x,y,z\}$ denote Cartesian directions,
$N_{\mathbf q}$ is the number of sampled wave vectors, and $V$ is the
primitive-cell volume. For phonon mode $\lambda=(\mathbf q,s)$, with wave
vector $\mathbf q$ and branch index $s$, $C_{\lambda}$ is the mode heat
capacity, $v_{\lambda,\alpha}$ is the $\alpha$ component of the group
velocity, and $F_{\lambda,\beta}$ is the $\beta$ component of the generalized
mean free displacement obtained from the direct LBTE solution. Conductivities in the single-mode relaxation-time approximation (RTA), which neglects the coupling between phonon modes retained in the direct solution,
were obtained from the same calculations for comparison. Further details of the LBTE formalism are
given in Text~S3.

The calculated states cover 0--\SI{135}{\giga\pascal} and
300--\SI{3800}{\kelvin} for \ce{MgO}, 0--\SI{135}{\giga\pascal} and
300--\SI{3500}{\kelvin} for Pv, and 100--\SI{145}{\giga\pascal} and
300--\SI{4000}{\kelvin} for PPv; the complete state list is given in
Table~S8. Reciprocal-space meshes of $24\times24\times24$,
$12\times12\times12$, and $14\times14\times14$ were used for \ce{MgO}, Pv,
and PPv, respectively, with Brillouin-zone integration by the
linear-tetrahedron method. Supercell and reciprocal-space convergence tests
are reported in Text~S3, Tables~S5--S6, and Fig.~S3; the resulting numerical
uncertainties are discussed in Section~\ref{sec:validation}.

Cubic symmetry makes the thermal-conductivity tensor of \ce{MgO} isotropic.
For orthorhombic Pv and PPv, the tensor is diagonal in the frame of the
lattice vectors, and its three components are compared in rank order as
$\kappa_{\min}$, $\kappa_{\mathrm{mid}}$, and $\kappa_{\max}$. For PPv,
these components are assigned to crystallographic directions in
Section~\ref{sec:silicates} and Text~S6. The orientational average is
defined as
\begin{equation}
\kappa_{\mathrm{latt}}
=
\frac{
\kappa_{\mathrm{latt},xx}
+
\kappa_{\mathrm{latt},yy}
+
\kappa_{\mathrm{latt},zz}
}{3}.
\label{eq:kappa_average}
\end{equation}
Throughout this paper, \kappalatt\ without Cartesian indices denotes the
isotropic conductivity of \ce{MgO} and the orientational average for Pv and
PPv.

Unless stated otherwise, \kappalatt\ denotes the intrinsic three-phonon
conductivity without isotope-disorder scattering. To quantify the effect of
natural isotopic disorder, the \ce{MgO} calculations at
\SI{0}{\giga\pascal} between \SI{300}{\kelvin} and
\SI{2000}{\kelvin} were repeated with phonon--isotope scattering treated in
second-order perturbation theory \citep{Tamura1983}, using the natural Mg and
O isotopic abundances tabulated in Phono3py. All other settings were held
fixed.

\subsection{Bridgmanite--post-perovskite phase boundary}
\label{sec:boundary}

The Pv--PPv phase boundary was determined from the DFT-QHA Gibbs free
energies of the two polymorphs (Eq.~\ref{eq:gibbs_qha}), normalized per
\ce{MgSiO3} formula unit. Both polymorphs were calculated using matched PAW
datasets and a common plane-wave cutoff of \SI{650}{\electronvolt}.
Brillouin-zone sampling was converged separately for each polymorph and held
fixed across its volume series. The common pressure grid and remaining
vibrational settings are reported in Text~S7.

The free-energy difference was defined as
$\Delta G(P,T)=G_{\mathrm{PPv}}(P,T)-G_{\mathrm{Pv}}(P,T)$, such that
$\Delta G<0$ denotes PPv stability. At each temperature, the transition
pressure was obtained by interpolating the zero crossing of $\Delta G(P,T)$.
The transition pressures between \SI{2000}{\kelvin} and
\SI{3500}{\kelvin} were fitted as
$P_{\mathrm{tr}}(T)=P_0+(\mathrm{d}P/\mathrm{d}T)_{\mathrm{tr}}T$, where
$(\mathrm{d}P/\mathrm{d}T)_{\mathrm{tr}}$ is the Clapeyron slope. The
intersection of the fitted boundary with the adopted geotherm was determined
by interpolation. Because this construction is quasi-harmonic, it neglects
intrinsic anharmonicity, which can curve the phase boundary at high
temperature \citep{ZhangWentzcovitch2022}.

\subsection{Density--temperature parameterization and aggregate conductivity}
\label{sec:fit}

For interpolation across lower-mantle conditions, the calculated
conductivities of each phase were fitted by least squares in logarithmic
space using the density--temperature relation of \citet{Haigis2012},
\begin{equation}
\kappa_{\mathrm{latt}}(\rho,T)
=
\kappa_{\mathrm{latt},0}
\left(
\frac{\rho}{\rho_0}
\right)^a
\left(
\frac{T_0}{T}
\right)^b,
\label{eq:kappa_parameterization}
\end{equation}
where $\rho$ is the density at the QHA equilibrium volume, $\rho_0$ and
$T_0$ are reference values, $\kappa_{\mathrm{latt},0}$ is the conductivity
at the reference state, and $a$ and $b$ are the fitted density and
temperature exponents. Only states at temperatures of
\SI{1500}{\kelvin} and above were included, over pressure ranges of
24--\SI{135}{\giga\pascal} for \ce{MgO} and Pv and
100--\SI{145}{\giga\pascal} for PPv. The parameterizations apply only within
these phase-specific fitting domains. Their parameters, uncertainties, and
residual statistics are reported in Text~S4 and Table~S7.

Along the lower-mantle geotherm of \citet{StaceyDavis2008}, the \ce{MgO}
parameterization was used throughout, while the Pv parameterization was used
at pressures below the calculated phase transition and the PPv
parameterization at higher pressures. Densities were obtained from the
quasi-harmonic thermal equations of state of the three phases (Table~S4).
The 95\% confidence intervals were obtained by propagating the parameter
covariances of the conductivity and thermal equation-of-state fits. They do
not include uncertainties in the adopted geotherm, the phase boundary, or
the underlying transport model; further details are given in Text~S4. The
directly calculated LBTE conductivities remain the primary results, and the
parameterizations are used only for interpolation.

For comparison with earlier aggregate estimates, the single-phase
conductivities along the geotherm were combined for an idealized Fe-free,
lattice-only aggregate containing 80~vol\% silicate (Pv or PPv) and
20~vol\% \ce{MgO}. The orientationally averaged conductivity was used for
each silicate phase. The Hashin--Shtrikman bounds
\citep{HashinShtrikman1962} for a two-phase isotropic composite were
calculated as
\begin{equation}
\kappa_{\mathrm{HS}}
=
\kappa_1
+
\frac{f_2}
{\left(\kappa_2-\kappa_1\right)^{-1}
+f_1/\left(3\kappa_1\right)},
\label{eq:hashin_shtrikman}
\end{equation}
where $f_i$ and $\kappa_i$ are the volume fraction and conductivity of phase
$i$, respectively. Equation~\ref{eq:hashin_shtrikman} was evaluated with the
more conductive phase as phase~1 for the upper bound and with the less
conductive phase as phase~1 for the lower bound. These are theoretical
microstructure bounds and not statistical confidence intervals.

\section{Results and discussion}
\label{sec:results}

\subsection{Validation of the fine-tuned MACE potentials}
\label{sec:validation}

\begin{figure*}[!tp]
  \centering
  \includegraphics[width=\textwidth]{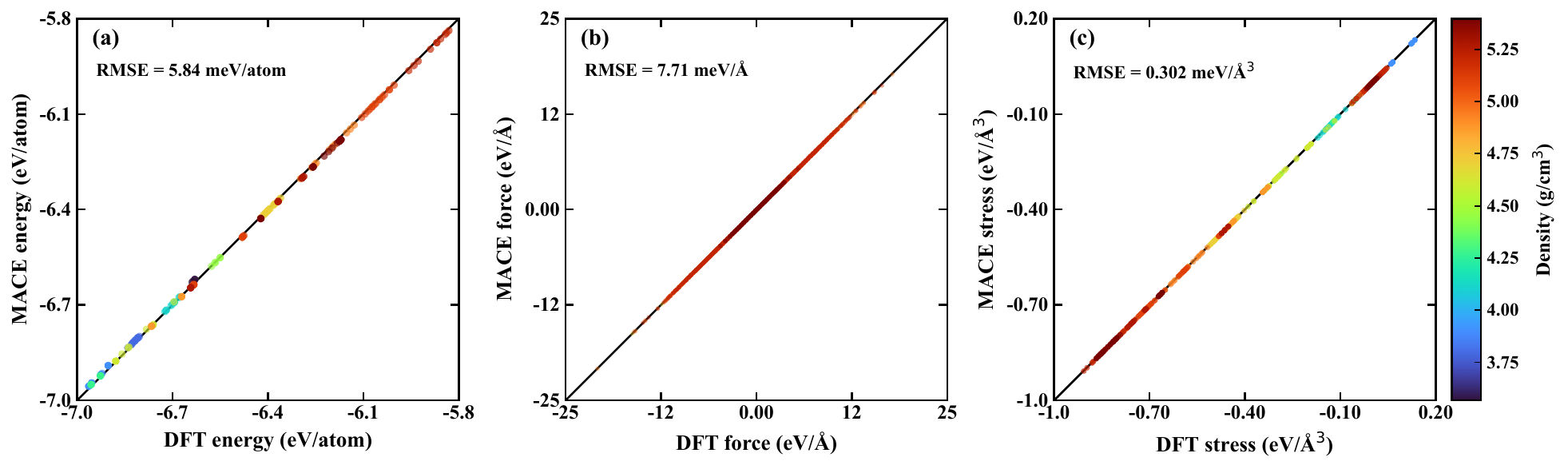}
  \caption{DFT--MACE parity plots for the held-out bridgmanite test set:
  (a) energy per atom, (b) force components, and (c) stress components.
  Points are coloured by density, and the black diagonal denotes exact
  agreement.}
  \label{fig:pv_parity}
\end{figure*}

On the independent test sets, the fine-tuned potentials reproduce the DFT
energies of \ce{MgO}, Pv, and PPv with root-mean-square errors (RMSEs) of
\SI{3.90}{\milli\electronvolt\per\atomunit},
\SI{5.84}{\milli\electronvolt\per\atomunit}, and
\SI{1.54}{\milli\electronvolt\per\atomunit}, the force components with
RMSEs of \SI{1.55}{\milli\electronvolt\per\angstromunit},
\SI{7.71}{\milli\electronvolt\per\angstromunit}, and
\SI{8.78}{\milli\electronvolt\per\angstromunit}, and the stress components
with RMSEs of \SI{0.226}{\milli\electronvolt\per\cubic\angstromunit},
\SI{0.302}{\milli\electronvolt\per\cubic\angstromunit}, and
\SI{0.393}{\milli\electronvolt\per\cubic\angstromunit}, respectively
(Table~S3). Relative to the RMS magnitudes of the DFT force components, the
force RMSEs are 0.18\%, 0.55\%, and 0.43\% for \ce{MgO}, Pv, and PPv,
respectively. The bridgmanite parity plot shows no systematic drift with density
over the sampled compression range (Fig.~\ref{fig:pv_parity}); corresponding
plots for \ce{MgO} and PPv are provided in Figs.~S1 and S2. Errors on
individual configurations, however, do not establish whether a potential
reproduces the curvature of the energy surface that enters lattice dynamics,
and we therefore tested the potentials against properties that were not
included directly in the fit.

\begin{figure}[!tbp]
  \centering
  \includegraphics[width=\columnwidth]{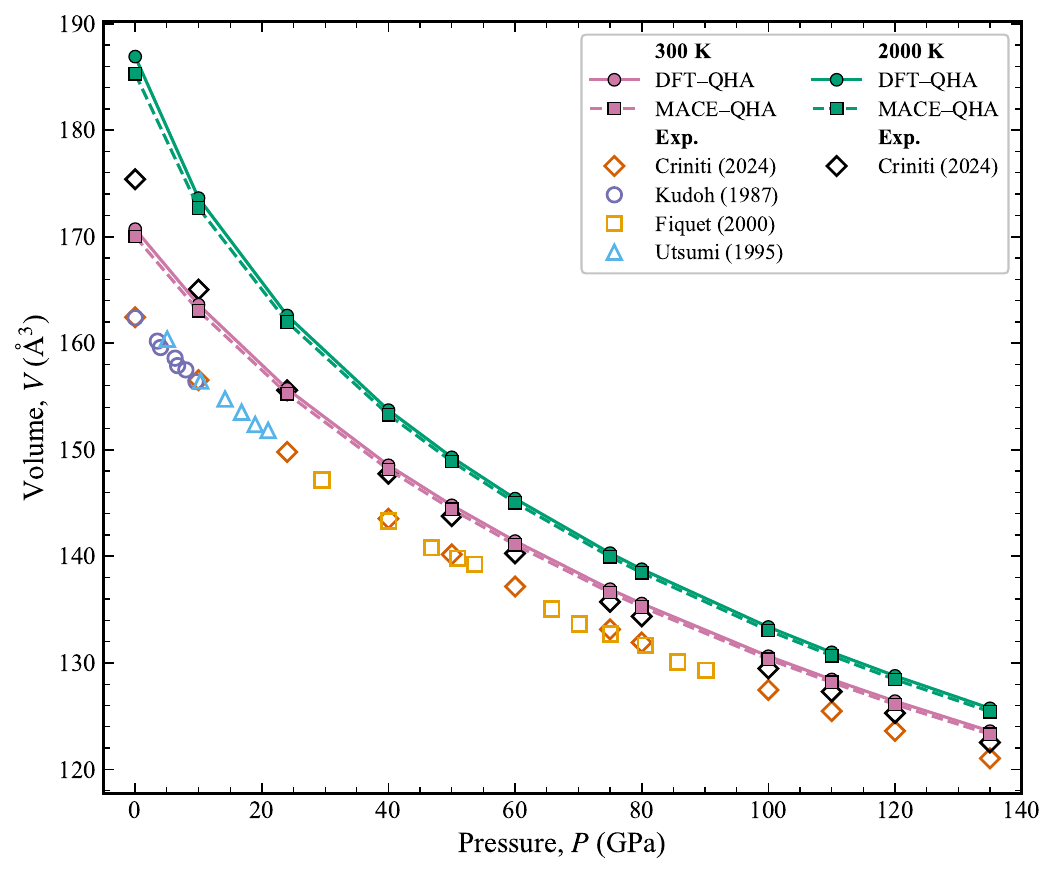}
  \caption{Quasi-harmonic pressure--volume relations of bridgmanite at
  300 and \SI{2000}{\kelvin}, comparing MACE-QHA and DFT-QHA calculations
  with the \ce{MgSiO3} end-member thermal equation of state of
  \citet{Criniti2024} and diffraction data from \citet{Kudoh1987},
  \citet{Utsumi1995}, and \citet{Fiquet2000}. Volumes are reported for the
  conventional four-formula-unit cell ($Z=4$).}
  \label{fig:eos_pv}
\end{figure}

The first such test is the quasi-harmonic equation of state. MACE-QHA and
DFT-QHA volumes were calculated independently, each using its corresponding
vibrational free energy. For Pv at pressures of \SI{24}{\giga\pascal} and
above, the MACE volumes are systematically smaller than the DFT volumes by
0.21--0.30\% at \SI{300}{\kelvin} and by 0.24--0.37\% at
\SI{2000}{\kelvin} (Fig.~\ref{fig:eos_pv}). The largest difference,
0.86\%, occurs at \SI{0}{\giga\pascal} and \SI{2000}{\kelvin}, outside the
stability field of bridgmanite and where the quasi-harmonic thermal expansion
is largest. The maximum absolute volume differences for \ce{MgO} and PPv are
0.35\% and 0.24\%, respectively (Text~S5 and Figs.~S4--S5). Agreement at
both 300 and \SI{2000}{\kelvin} indicates that the potentials reproduce not
only static compression but also the volume dependence of the phonon
frequencies that controls quasi-harmonic thermal expansion.

The MACE--DFT volume differences are substantially smaller than the
systematic offset between the PBE reference calculations and experimental
thermoelastic data. At \SI{300}{\kelvin}, the DFT-QHA Pv volumes exceed the
best-fit \ce{MgSiO3} end-member thermal equation of state reported by
\citet{Criniti2024} by 5.1\% at \SI{0}{\giga\pascal} and 2.1\% at
\SI{135}{\giga\pascal}, where that curve is extrapolated, whereas the
room-temperature diffraction data of \citet{Kudoh1987},
\citet{Utsumi1995}, and \citet{Fiquet2000} lie within 0.7\% of it
(Fig.~\ref{fig:eos_pv}). The larger DFT volumes are consistent with the
underbinding tendency of PBE. Because the transport calculations used the
DFT-QHA volumes, this density offset is inherited from the reference
functional and is not introduced by fine-tuning. Given the positive density
dependence of \kappalatt, it also tends, all else being equal, to lower the
calculated conductivity at a specified pressure relative to a calculation at
the experimental density.

\begin{figure*}[!tp]
  \centering
  \includegraphics[width=0.49\textwidth]{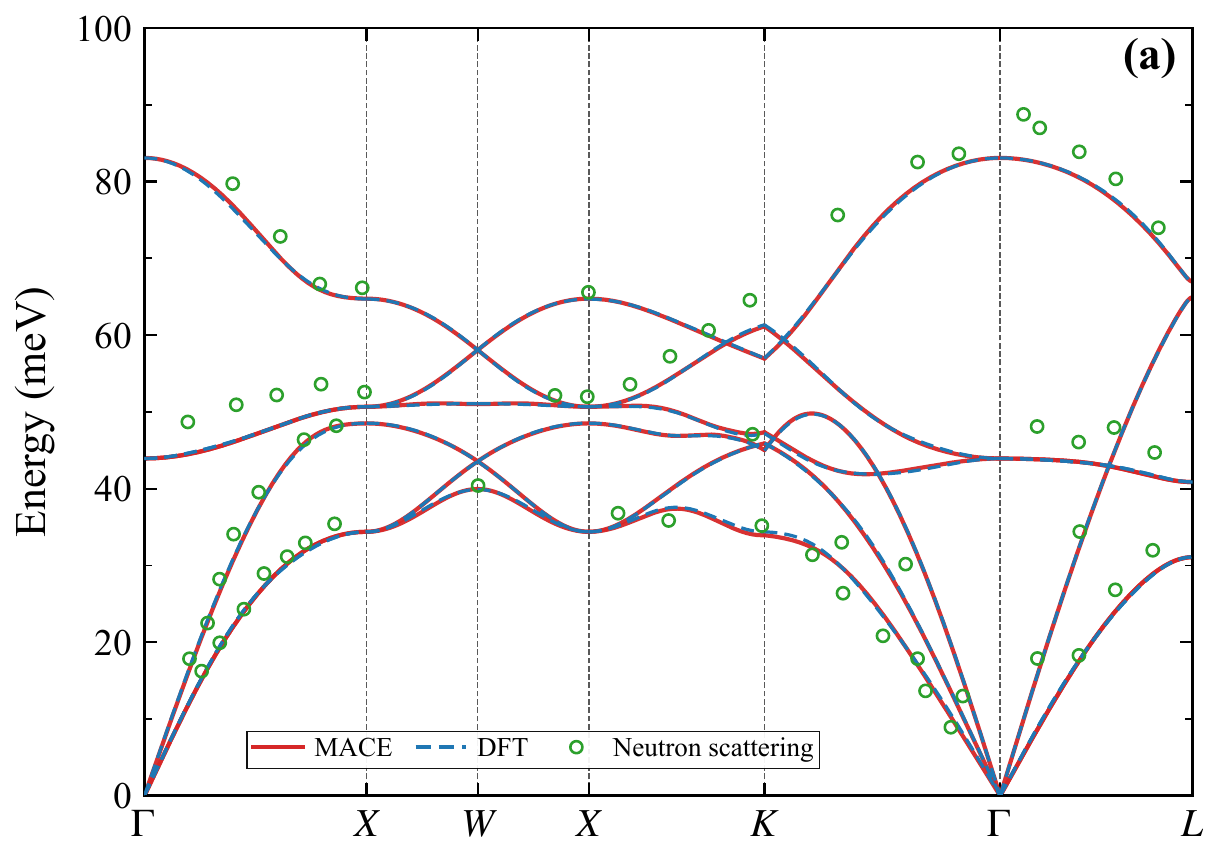}\hfill
  \includegraphics[width=0.49\textwidth]{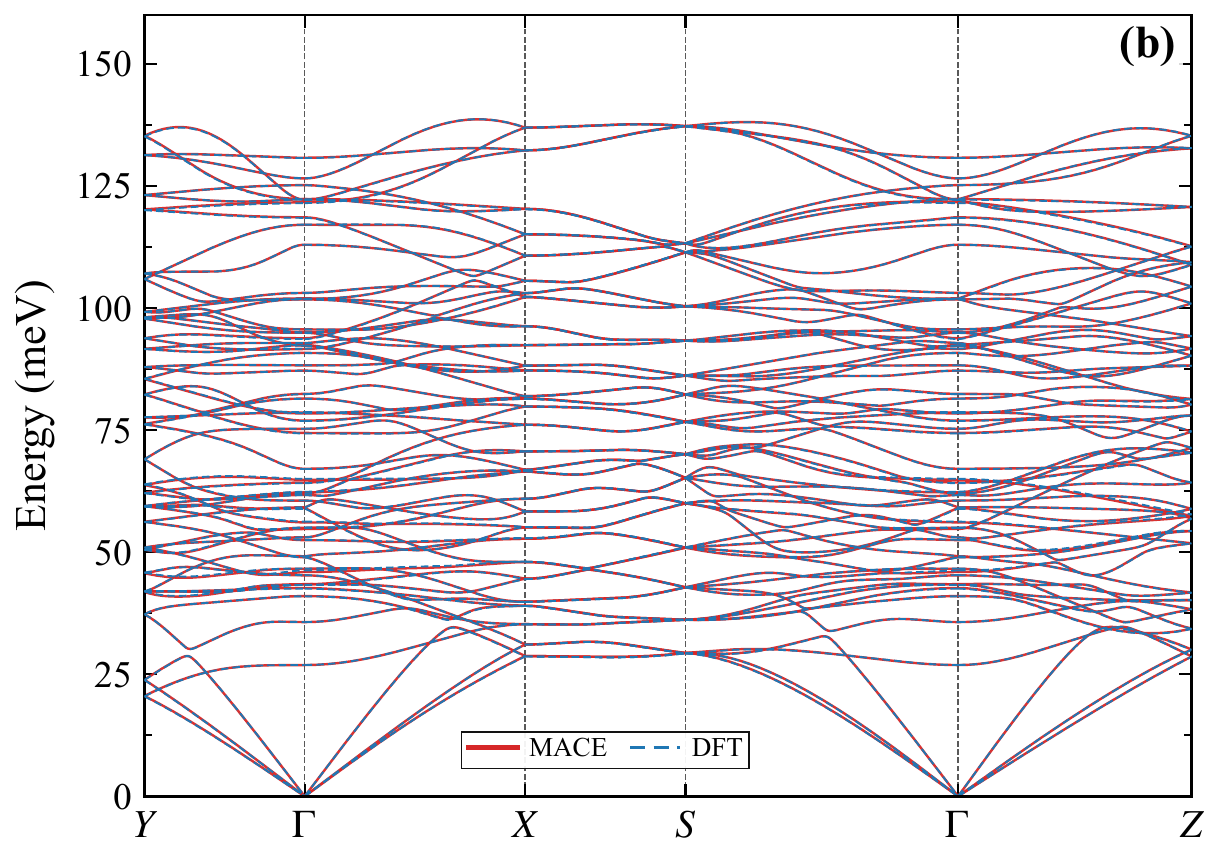}
  \caption{Phonon dispersions obtained from MACE and DFT force constants for
  (a) \ce{MgO} at \SI{0}{\giga\pascal} and \SI{300}{\kelvin}, compared with
  the inelastic neutron-scattering data of \citet{Sangster1970}, and
  (b) bridgmanite at \SI{100}{\giga\pascal} and
  \SI{3000}{\kelvin}. The calculations were performed at the corresponding
  QHA equilibrium structures. Frequencies are plotted in meV
  (\SI{1}{\tera\hertz} = \SI{4.136}{\milli\electronvolt}). An additional
  ambient-condition bridgmanite comparison with the inelastic
  X-ray-scattering data of \citet{Wehinger2016}, together with the
  compressed-state post-perovskite and \ce{MgO} comparisons, is shown in
  Fig.~S6.}
  \label{fig:phonons_main}
\end{figure*}

Phonon dispersions provide a second test of the potentials against a property
not fitted directly. At the compressed QHA equilibrium structures of
\ce{MgO} at \SI{120}{\giga\pascal} and \SI{3200}{\kelvin}, Pv at
\SI{100}{\giga\pascal} and \SI{3000}{\kelvin}, and PPv at
\SI{135}{\giga\pascal} and \SI{3500}{\kelvin}, the RMS frequency
differences between MACE and DFT, evaluated over all phonon branches and
sampled wave vectors, are 0.008, 0.016, and
\SI{0.014}{\tera\hertz}, respectively. Across the three comparisons, the
maximum absolute difference for any frequency at a sampled wave vector is
\SI{0.083}{\tera\hertz}. The compressed-state comparisons are shown for
\ce{MgO} in Fig.~S6c, for Pv in Fig.~\ref{fig:phonons_main}b, and for PPv
in Fig.~S6b. No unstable branches occur in either the MACE or DFT
dispersions. Ambient-condition comparisons with experimental data are shown
for \ce{MgO} in Fig.~\ref{fig:phonons_main}a \citep{Sangster1970} and for
Pv in Fig.~S6a \citep{Wehinger2016}. Because the harmonic force constants
determine the heat capacities, group velocities, and phase space available
for three-phonon scattering, this agreement is a necessary condition for
using the potentials in transport calculations.

The most direct test propagates the force differences through the complete
transport calculation. For \ce{MgO} at \SI{0}{\giga\pascal}, recalculating
the force constants with DFT forces on the same displaced structures, while
keeping the supercell, reciprocal-space mesh, and LBTE solver unchanged,
gives \kappalatt\ values of 49.05 and
\SI{3.44}{\watt\per\metre\per\kelvin} at 300 and
\SI{2000}{\kelvin}, respectively, compared with 49.16 and
\SI{3.37}{\watt\per\metre\per\kelvin} from MACE. Relative to DFT, the MACE
values differ by $+0.23\%$ and $-1.9\%$, respectively, isolating the effect
of replacing direct DFT force evaluations with the fine-tuned potential.
These differences are well below the spread of approximately $\pm15\%$
found when independent groups applied complete first-principles workflows to
the same benchmark systems \citep{McGaughey2025}. Corresponding direct DFT
transport calculations were not performed for Pv or PPv: a single
thermodynamic state would require DFT forces for 14\,675 or 15\,440 displaced
supercells containing 160 or 320 atoms, respectively. For the silicates,
validation therefore rests on the held-out test sets, equations of state,
and phonon dispersions above, together with comparisons against published
first-principles LBTE results discussed below.

The remaining numerical sensitivity was assessed by increasing the
supercell sizes and refining the reciprocal-space meshes. Increasing the
supercell size changed \kappalatt\ by 2.3--2.8\% for \ce{MgO}, 0.23\% for
Pv, and 0.12\% for PPv. Refining the reciprocal-space meshes changed
\kappalatt\ by at most 0.13\% for \ce{MgO}, by 2.0--2.4\% for Pv, and by
0.23\% for PPv (Text~S3, Tables~S5--S6, and Fig.~S3). For \ce{MgO}, the
residual finite-supercell effect is therefore larger than the difference
caused by replacing DFT forces with MACE forces. These comparisons apply
within the quasi-harmonic, three-phonon LBTE framework used here. Where a
direct transport-level benchmark was feasible, replacing DFT forces with the
fine-tuned potential introduced a discrepancy smaller than the residual
numerical sensitivity of the transport calculation.

\subsection{Lattice thermal conductivity of MgO}
\label{sec:mgo}

\begin{figure}[!tbp]
  \centering
  \includegraphics[width=\columnwidth]
  {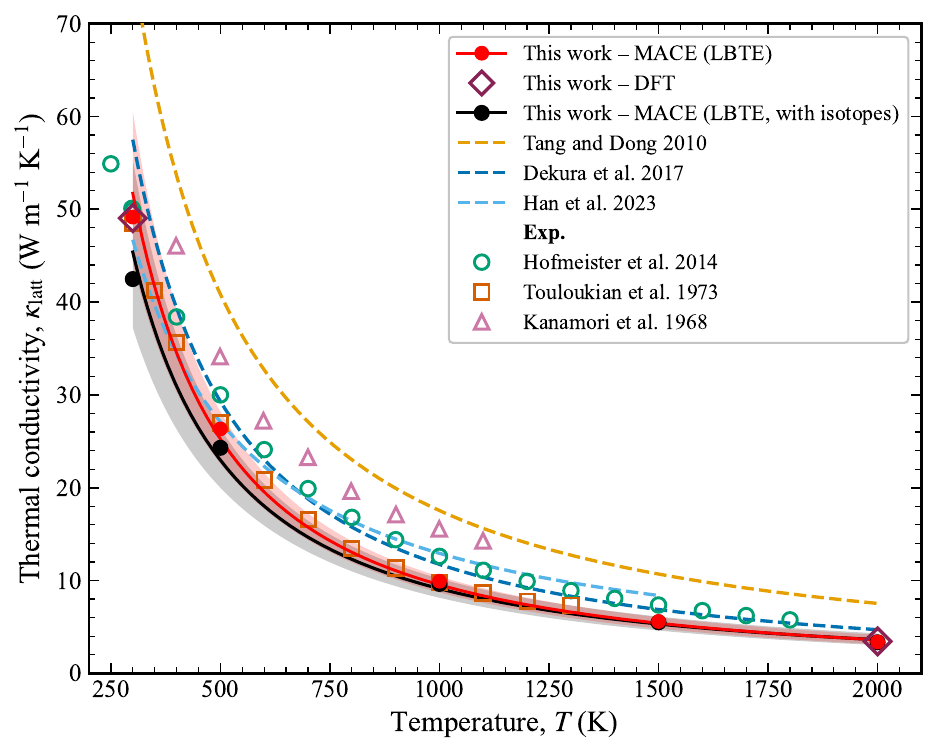}
  \caption{Temperature dependence of the lattice thermal conductivity of
  \ce{MgO} at \SI{0}{\giga\pascal}. Full-LBTE results obtained with MACE
  forces are shown with and without natural-isotope scattering, together
  with direct DFT-force benchmarks at 300 and \SI{2000}{\kelvin}. Solid
  curves are power-law fits to the present MACE results, with shaded regions
  indicating 95\% regression confidence intervals. Dashed curves show the
  calculations of \citet{TangDong2010} and
  \citet{DekuraTsuchiya2017}, and the phonon-renormalized calculation
  including three- and four-phonon scattering of \citet{Han2023MgO}. Open
  symbols show the measurements of \citet{Kanamori1968},
  \citet{Touloukian1973}, and \citet{Hofmeister2014}.}
  \label{fig:mgo_temperature}
\end{figure}

At zero pressure, the lattice thermal conductivity of \ce{MgO} decreases from
\SI{49.2}{\watt\per\metre\per\kelvin} at \SI{300}{\kelvin} to 26.3, 9.9,
5.6, and \SI{3.4}{\watt\per\metre\per\kelvin} at 500, 1000, 1500, and
\SI{2000}{\kelvin}, respectively
(Fig.~\ref{fig:mgo_temperature}). A power-law fit over this interval,
$\kappalatt\propto T^{-n}$, gives $n=1.40$, indicating a steeper decrease
than the $T^{-1}$ dependence expected for three-phonon scattering in the
classical limit at fixed volume. Thermal expansion contributes to this
difference: the QHA density decreases by 7.0\% between 300 and
\SI{2000}{\kelvin}. At fixed density, by contrast, the lower-mantle
parameterization described below gives an approximately $T^{-1}$ dependence.
The isotope-free ambient value of
\SI{49.2}{\watt\per\metre\per\kelvin} is about 8\% below the LDA-based
full-LBTE result of \SI{53.7}{\watt\per\metre\per\kelvin} reported by
\citet{DekuraTsuchiya2017}, in the direction expected from the larger PBE
equilibrium volume discussed in Section~\ref{sec:validation}.

Natural isotopic disorder lowers the calculated ambient conductivity to
\SI{42.5}{\watt\per\metre\per\kelvin} at \SI{300}{\kelvin}, a reduction of
14\%. As anharmonic scattering strengthens with temperature, the relative
isotope effect decreases monotonically, reaching 1\% at
\SI{2000}{\kelvin}. Expressed relative to the isotope-included value, the
isotope-free conductivity is 16\% higher, compared with the increase of up
to 46\% reported by \citet{TangDong2010} and subsequently cited by
\citet{Dalton2013}. \citet{DekuraTsuchiya2017} showed that restricting the
third-order interactions to second-nearest neighbours, as in the calculations
of \citet{TangDong2010}, substantially overestimates the conductivity. Weaker
intrinsic three-phonon scattering correspondingly increases the relative
contribution of isotope scattering. The isotope-included conductivities lie
close to the values compiled by \citet{Touloukian1973} but below the
single-crystal laser-flash data of \citet{Hofmeister2014}, by about 15\% at
\SI{300}{\kelvin} and 24--25\% between 1000 and
\SI{1500}{\kelvin}. The offset at \SI{300}{\kelvin} is consistent with the
4.5\% larger PBE volume relative to experiment \citep{Speziale2001}: with
the density exponents of 4.4--5.2 found at high pressure, a volume difference
of this magnitude would lower \kappalatt\ by approximately 17--20\%. At
higher temperatures, the differences among the measurements themselves are
comparable to this offset (Fig.~\ref{fig:mgo_temperature}). Because isotope
scattering was evaluated only at zero pressure, the resulting correction has
not been extrapolated to lower-mantle pressures.

Direct solution of the LBTE gives systematically higher conductivities than
the relaxation-time approximation (RTA). At zero pressure, the full solution
exceeds the RTA by 12.7\% at \SI{300}{\kelvin} and 5.5\% at
\SI{2000}{\kelvin}; at \SI{135}{\giga\pascal}, the difference reaches
24\% at both 300 and \SI{3500}{\kelvin}, indicating a stronger correction
under compression. \citet{DekuraTsuchiya2017} likewise obtained full-LBTE
values 27\% and 26\% above their RTA results at ambient and core--mantle
boundary conditions, respectively:
\SI{53.7}{\watt\per\metre\per\kelvin} versus
\SI{42.2}{\watt\per\metre\per\kelvin}, and
\SI{39.7}{\watt\per\metre\per\kelvin} versus
\SI{31.6}{\watt\per\metre\per\kelvin}.

\begin{figure*}[!tp]
  \centering
  \includegraphics[width=0.86\textwidth]{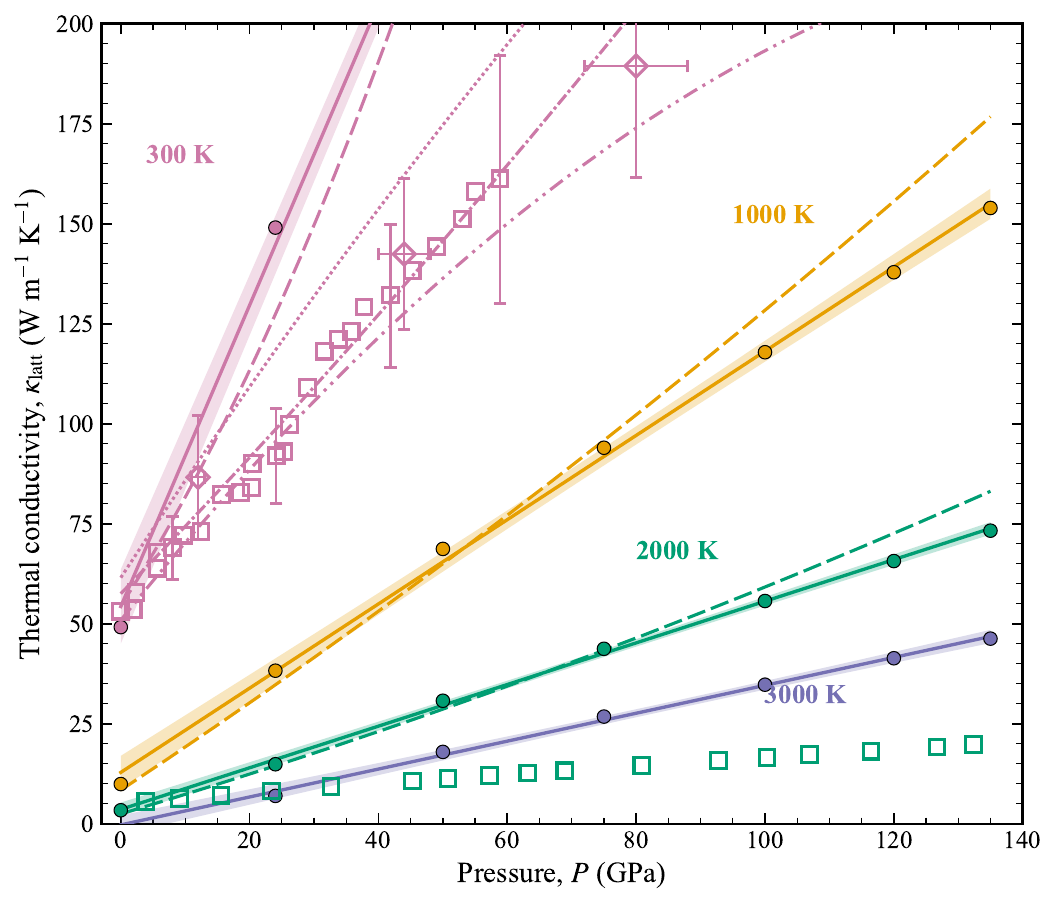}
  \caption{Pressure dependence of the lattice thermal conductivity of
  \ce{MgO} at 300, 1000, 2000, and \SI{3000}{\kelvin}. Colours identify
  temperature. Filled circles show the present direct-LBTE calculations;
  solid lines are linear fits, with shaded regions indicating 95\%
  regression confidence intervals. Literature calculations are shown over
  their supported pressure ranges: dashed, \citet{DekuraTsuchiya2017};
  dotted, \citet{TangDong2010}; dash-dotted, \citet{Stackhouse2010}; and
  dash-dot-dot, \citet{deKoker2010}. Open squares denote the 300~K
  measurements and 2000~K model values of \citet{Dalton2013}; open diamonds
  denote the 300~K measurements of \citet{Hasegawa2023}. The vertical axis
  is limited to \SI{200}{\watt\per\metre\per\kelvin}; the 300~K calculations
  above \SI{24}{\giga\pascal} and the \SI{130}{\giga\pascal} measurement of
  \citet{Hasegawa2023} lie above the plotted range.}
  \label{fig:mgo_pressure}
\end{figure*}

Compression strongly increases \kappalatt. At \SI{2000}{\kelvin}, the
conductivity rises from \SI{14.9}{\watt\per\metre\per\kelvin} at
\SI{24}{\giga\pascal} to \SI{73.3}{\watt\per\metre\per\kelvin} at
\SI{135}{\giga\pascal}. At \SI{3000}{\kelvin}, it increases from 7.0 to
\SI{46.3}{\watt\per\metre\per\kelvin} over the same pressure interval
(Fig.~\ref{fig:mgo_pressure}). The 32 states between 24 and
\SI{135}{\giga\pascal} and between 1500 and \SI{3800}{\kelvin} are described
by Eq.~\ref{eq:kappa_parameterization} with $a=5.20\pm0.13$,
$b=1.01\pm0.05$, and a mean absolute percentage error of 6.0\%
(Table~S7). The fitted density exponent is consistent with $4.6\pm1.2$ from
first-principles nonequilibrium molecular dynamics \citep{Stackhouse2010}
and close to the value of 5.42 from classical Green--Kubo simulations
\citep{Haigis2012}, but higher than $4.4\pm0.1$ from high-pressure
thermoreflectance measurements \citep{Hasegawa2023}. The temperature
exponents show greater variation: classical simulations give 1.10
\citep{Haigis2012}, whereas experimental parameterizations give 0.62 for
single-crystal \ce{MgO} \citep{Hasegawa2023} and 0.76 for polycrystalline
material \citep{Manthilake2011}.

Expressing the results in terms of density permits a direct comparison with
high-temperature measurements without retaining the PBE pressure--density
offset. \citet{Hasegawa2023} measured single-crystal \ce{MgO} to about
\SI{140}{\giga\pascal} at \SI{1950}{\kelvin} and to
\SI{3170}{\kelvin} at lower pressures, reporting the density of each state
from an experimental equation of state. Evaluated at those densities and
temperatures, the present parameterization agrees with all nine
high-pressure, high-temperature measurements within $2\sigma$, including
seven within $1\sigma$. For example, the measured value of
$88.8\pm19.9~\si{\watt\per\metre\per\kelvin}$ at
\SI{140}{\giga\pascal} and \SI{1950}{\kelvin} compares with a calculated
value of \SI{84.3}{\watt\per\metre\per\kelvin}, while measured values of
7.3--\SI{8.0}{\watt\per\metre\per\kelvin} near
\SI{30}{\giga\pascal} and \SI{3000}{\kelvin} compare with calculated values
of 6.8--\SI{7.8}{\watt\per\metre\per\kelvin}. The other two measurements,
at 19 and \SI{57}{\giga\pascal}, differ from the calculated values by
$1.5\sigma$ and $1.1\sigma$, respectively. Five experimental densities
lie just outside the fitted interval, by at most 3.3\%, and two temperatures
lie below its \SI{1500}{\kelvin} lower limit; these comparisons therefore
involve only short extrapolations.

At \SI{300}{\kelvin}, by contrast, the calculated conductivities exceed the
high-pressure measurements. The calculated value of
\SI{567}{\watt\per\metre\per\kelvin} at
\SI{135}{\giga\pascal} is approximately twice the single-crystal value of
$275\pm64~\si{\watt\per\metre\per\kelvin}$ measured at
\SI{130}{\giga\pascal} \citep{Hasegawa2023}, and the discrepancy is larger
relative to the polycrystalline measurements of \citet{Imada2014}. The
logarithmic pressure slope calculated between 0 and
\SI{24}{\giga\pascal} is \SI{4.6}{\percent\per\giga\pascal} on average,
compared with an initial value of
\SI{1.9}{\percent\per\giga\pascal} measured by \citet{Dalton2013}. At room
temperature, long phonon mean free paths make the measured conductivity
particularly sensitive to extrinsic scattering. The high-pressure
calculations omit isotopic disorder, point defects, dislocations, and grain
boundaries; the separate isotope test applies only at zero pressure. The
high-temperature agreement is therefore the more relevant test of the
intrinsic transport calculation for lower-mantle conditions.

At the CMB point of the adopted geotherm
(\SI{135.8}{\giga\pascal}, \SI{3739}{\kelvin}), the parameterization gives
\SI{37.3}{\watt\per\metre\per\kelvin}. At
\SI{135}{\giga\pascal} and \SI{3800}{\kelvin}, the directly calculated value
of \SI{35.1}{\watt\per\metre\per\kelvin} is about 12\% below the full-LBTE
result of \SI{39.7}{\watt\per\metre\per\kelvin} obtained by
\citet{DekuraTsuchiya2017} at \SI{136}{\giga\pascal} and the same
temperature, in the direction expected from the larger PBE volume. At
\SI{135}{\giga\pascal} and \SI{3000}{\kelvin}, the present value of
\SI{46.3}{\watt\per\metre\per\kelvin} is also close to the approximately
\SI{43}{\watt\per\metre\per\kelvin} reported by
\citet{TangDong2010}. Earlier theoretical estimates for comparable
lowermost-mantle conditions span approximately
\SIrange{20}{81}{\watt\per\metre\per\kelvin}
\citep{Stackhouse2010,deKoker2010,Haigis2012,DekuraTsuchiya2017}.
Experimentally based estimates of
$17.9\pm1.1~\si{\watt\per\metre\per\kelvin}$ \citep{Imada2014} and
$51.8^{+10.5}_{-8.7}~\si{\watt\per\metre\per\kelvin}$
\citep{Hasegawa2023} are extrapolations from \SI{300}{\kelvin} and
\SI{1950}{\kelvin}, using temperature exponents of 0.76 and 0.62,
respectively.

For the intrinsic \ce{MgO} conductivity, the largest remaining physical
uncertainty is the treatment of anharmonicity beyond three-phonon scattering
at quasi-harmonic volumes. Finite-temperature phonon renormalization would
require temperature-dependent effective phonons at each
pressure--temperature state, whereas four-phonon scattering would require
fourth-order force constants and a higher-order transport implementation,
each with its own convergence tests. Retaining the same three-phonon
framework for all three phases keeps the results directly comparable with
previous direct-LBTE calculations and with the DFT-force benchmark in
Section~\ref{sec:validation}. Including both effects, \citet{Chabane2025}
obtained \SI{50.7}{\watt\per\metre\per\kelvin} at
\SI{135}{\giga\pascal} and \SI{3570}{\kelvin}, whereas the present
parameterization gives approximately
\SI{38}{\watt\per\metre\per\kelvin} at the same conditions. This difference
of about 25\% lies within the 20--45\% underestimate that those authors found
for a three-phonon-only description along the lower-mantle geotherm. The
\ce{MgO} conductivities reported here are therefore numerically converged
values within the quasi-harmonic, three-phonon framework, not evidence that
higher-order anharmonic effects are negligible.

\subsection{Lattice thermal conductivity of bridgmanite and post-perovskite}
\label{sec:silicates}

\begin{figure*}[!tp]
  \centering
  \includegraphics[width=0.495\textwidth]{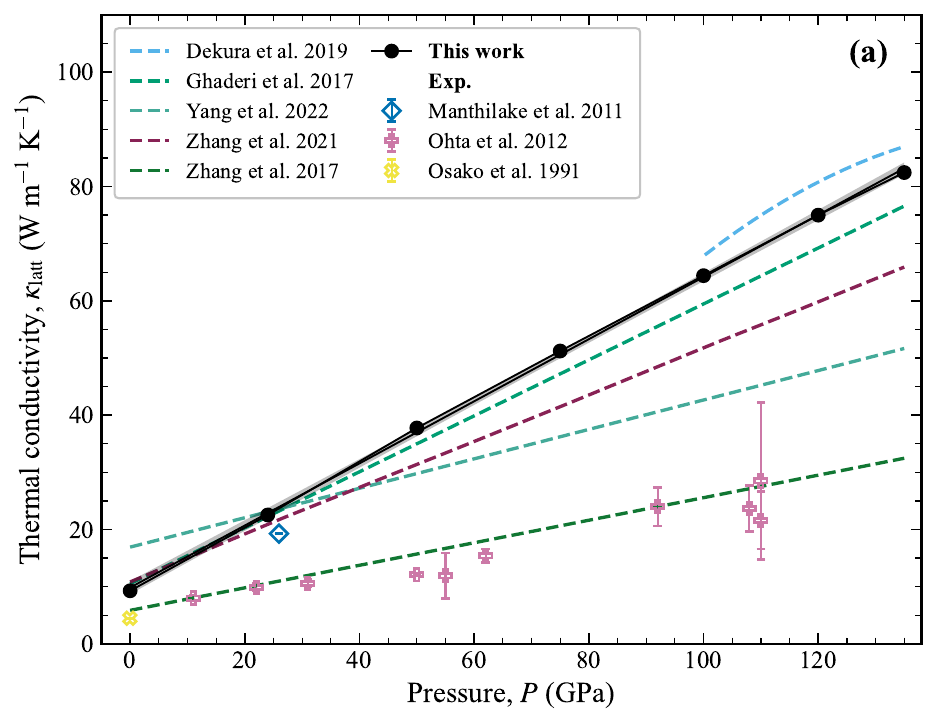}\hfill
  \includegraphics[width=0.495\textwidth]{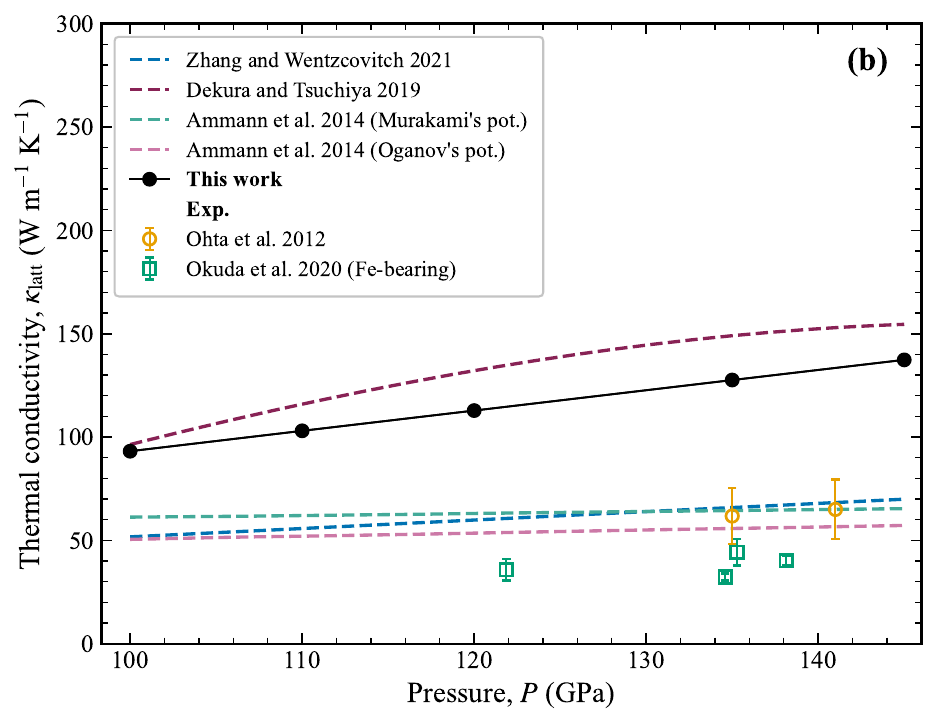}
  \caption{Orientationally averaged lattice thermal conductivity of
  (a) bridgmanite and (b) post-perovskite at \SI{300}{\kelvin}. Bridgmanite
  values are compared with the calculations of 
  \citet{Ghaderi2017}, \citet{Zhang2017}, \citet{DekuraTsuchiya2019},
  \citet{ZhangWentzcovitch2021}, and \citet{Yang2022}, and the measurements
  of \citet{Osako1991} and \citet{Ohta2012}; the value from
  \citet{Manthilake2011} is extrapolated to 300~K using their fitted
  temperature dependence.
  Post-perovskite values are compared with the calculations of
  \citet{Ammann2014}, \citet{DekuraTsuchiya2019}, and
  \citet{ZhangWentzcovitch2021}, and the measurements of \citet{Ohta2012}
  and \citet{Okuda2020}; the latter are for an Fe-bearing sample. Present
  values are intrinsic three-phonon conductivities without isotope
  scattering.}
  \label{fig:silicates_average}
\end{figure*}

At \SI{300}{\kelvin}, the orientationally averaged lattice thermal
conductivity of Pv increases from
\SI{9.3}{\watt\per\metre\per\kelvin} at \SI{0}{\giga\pascal} to 22.5,
64.4, and \SI{82.4}{\watt\per\metre\per\kelvin} at 24, 100, and
\SI{135}{\giga\pascal}, respectively
(Fig.~\ref{fig:silicates_average}a). At \SI{100}{\giga\pascal}, the present
value is 9\% above the full-LBTE result of
\SI{59.2}{\watt\per\metre\per\kelvin} reported by
\citet{Ghaderi2017}. The value at \SI{24}{\giga\pascal} is also close to
the \SI{23.6}{\watt\per\metre\per\kelvin} obtained at
\SI{26}{\giga\pascal} from deep-potential Green--Kubo simulations
\citep{Yang2022}. By comparison, the relaxation-time calculations of
\citet{Dekura2013} gave 9.8 and
\SI{28.1}{\watt\per\metre\per\kelvin} at 23.5 and
\SI{100}{\giga\pascal}, substantially below the full-LBTE results.
Room-temperature measurements on polycrystalline samples also lie below the
intrinsic calculations, from \SI{5.1}{\watt\per\metre\per\kelvin} at
ambient pressure \citep{Osako1991} to the high-pressure measurements of
\citet{Ohta2012}. Correcting for the PBE volume overestimate would raise the
calculated conductivity further, so this discrepancy cannot arise from the
density offset; as in the high-pressure room-temperature \ce{MgO}
comparison, extrinsic scattering in the samples is the more likely cause.
Natural isotopes, which are not included in the present silicate
calculations, were estimated by \citet{Ghaderi2017} to lower the Pv
conductivity by only 1--3\%.

\begin{figure*}[!tp]
  \centering
  \includegraphics[width=0.84\textwidth]
  {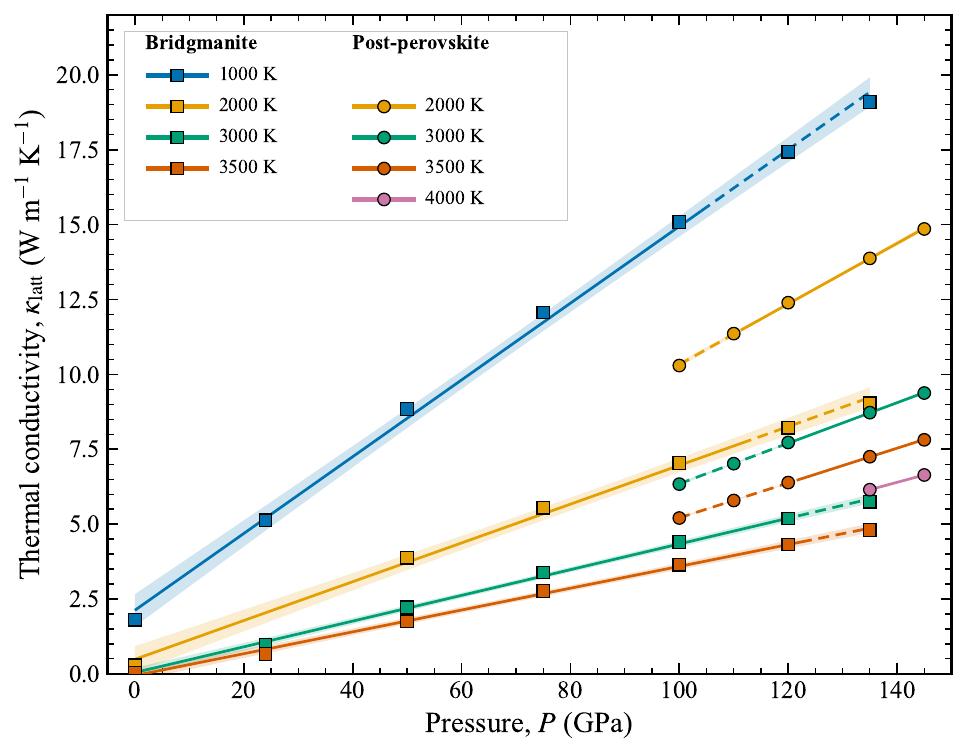}
  \caption{Pressure dependence of the orientationally averaged lattice
  thermal conductivity of bridgmanite (squares) at 1000, 2000, 3000, and
  \SI{3500}{\kelvin} and post-perovskite (circles) at 2000, 3000, 3500,
  and \SI{4000}{\kelvin}. Filled symbols show the direct-LBTE calculations,
  lines are linear fits, and shaded regions denote 95\% regression
  confidence bands. Solid line segments denote the stable phase according
  to the calculated Pv--PPv boundary, and dashed segments its metastable
  continuation}
  \label{fig:silicates_high_temperature}
\end{figure*}

At high temperature, the Pv conductivity is much lower than at
\SI{300}{\kelvin} but remains strongly pressure dependent. At
\SI{135}{\giga\pascal}, \kappalatt\ decreases from
\SI{9.0}{\watt\per\metre\per\kelvin} at \SI{2000}{\kelvin} to 5.7 and
\SI{4.8}{\watt\per\metre\per\kelvin} at 3000 and
\SI{3500}{\kelvin}, respectively
(Fig.~\ref{fig:silicates_high_temperature}). The
\SI{2000}{\kelvin} value is 9\% below the full-LBTE result of
\citet{DekuraTsuchiya2019} at \SI{136}{\giga\pascal}. The results over
24--\SI{135}{\giga\pascal} and 1500--\SI{3500}{\kelvin} are described by
Eq.~\ref{eq:kappa_parameterization} with $a=5.61\pm0.14$ and
$b=0.97\pm0.04$ (Table~S7). The temperature exponent at fixed density is
consistent with the value of 1.01 obtained from phonon-quasiparticle
calculations \citep{ZhangWentzcovitch2021}, but larger than the value of
0.61 obtained from machine-learning Green--Kubo simulations
\citep{PengDeng2024}. It is also larger than the exponent of 0.43 fitted by
\citet{Manthilake2011} and the values of approximately 0.2--0.5 obtained by
\citet{Edmund2024} across several compositions, although the latter carry
substantial uncertainties.

The difference in temperature dependence is evident at moderate
temperature. The fit of \citet{Manthilake2011} to polycrystalline
measurements gives a reference conductivity of
\SI{12.8}{\watt\per\metre\per\kelvin} at
\SI{26}{\giga\pascal} and \SI{700}{\kelvin}, whereas interpolation of the
present results at \SI{24}{\giga\pascal} gives approximately
\SI{8}{\watt\per\metre\per\kelvin} at the same temperature. Although the
PBE density offset contributes to this difference, it lowers the calculated
values at all temperatures alike; the calculated near-$T^{-1}$ dependence
is therefore appreciably stronger than the experimental trend, as was also
found for \ce{MgO}. This difference may partly reflect
temperature-independent scattering by grain boundaries and defects, which
weakens the apparent temperature dependence of real samples, as well as
anharmonic effects beyond three-phonon scattering that are absent from the
present calculations. For context, previous lowermost-mantle estimates range
from 1--\SI{2}{\watt\per\metre\per\kelvin} \citep{Tang2014} to
$6.8\pm0.9~\si{\watt\per\metre\per\kelvin}$ at
\SI{136}{\giga\pascal} and \SI{4000}{\kelvin}
\citep{Stackhouse2015}. The present Pv value of
\SI{4.8}{\watt\per\metre\per\kelvin} at
\SI{135}{\giga\pascal} and \SI{3500}{\kelvin} lies between these estimates,
although the thermodynamic conditions are not identical.

PPv is more conductive than Pv at every pressure--temperature state for
which the two phases were evaluated. At \SI{300}{\kelvin}, its
orientationally averaged conductivity increases from
\SI{93.1}{\watt\per\metre\per\kelvin} at
\SI{100}{\giga\pascal} to \SI{137}{\watt\per\metre\per\kelvin} at
\SI{145}{\giga\pascal}. At \SI{135}{\giga\pascal}, it decreases from
\SI{13.9}{\watt\per\metre\per\kelvin} at \SI{2000}{\kelvin} to 8.7 and
\SI{6.2}{\watt\per\metre\per\kelvin} at 3000 and
\SI{4000}{\kelvin}, respectively
(Fig.~\ref{fig:silicates_high_temperature}). The
\SI{2000}{\kelvin} value is 11\% below the full-LBTE result of
\SI{15.6}{\watt\per\metre\per\kelvin} reported by
\citet{DekuraTsuchiya2019} at \SI{136}{\giga\pascal}, whereas the
\SI{4000}{\kelvin} value agrees with their result of
\SI{6.2}{\watt\per\metre\per\kelvin}. The results over
100--\SI{145}{\giga\pascal} and 1500--\SI{4000}{\kelvin} are described by
Eq.~\ref{eq:kappa_parameterization} with $a=4.99\pm0.04$ and
$b=1.005\pm0.003$ (Table~S7).

The phonon-quasiparticle calculations of
\citet{ZhangWentzcovitch2021} predict a weaker temperature dependence,
$T^{-0.85}$, and a lower PPv conductivity. At
\SI{5.41}{\gram\per\cubic\centi\metre} and \SI{2000}{\kelvin}, their
parameterization gives \SI{9.4}{\watt\per\metre\per\kelvin}, compared with
\SI{14.3}{\watt\per\metre\per\kelvin} from the present parameterization,
a difference of 34\%. Measurements of Fe-bearing PPv at
\SI{138}{\giga\pascal} and temperatures up to \SI{1558}{\kelvin} likewise
give a weaker temperature dependence, with an exponent of $0.65\pm0.04$
\citep{Okuda2020}. At room temperature, the present PPv value of
\SI{128}{\watt\per\metre\per\kelvin} at
\SI{135}{\giga\pascal} is lower than the classical Green--Kubo value of
$167\pm25~\si{\watt\per\metre\per\kelvin}$ at
\SI{138}{\giga\pascal} and \SI{298}{\kelvin} \citep{Haigis2012}, but is
approximately twice the measured value of
$61.8\pm13.6~\si{\watt\per\metre\per\kelvin}$ at
\SI{135}{\giga\pascal} and \SI{300}{\kelvin} \citep{Ohta2012}. The
measurement of \citet{Ohta2012} therefore lies well below the intrinsic
calculation, as do the high-pressure room-temperature measurements for
\ce{MgO} and Pv, consistent with additional extrinsic scattering in the
experimental samples (Fig.~\ref{fig:silicates_average}b).

Compared at the same pressure and temperature, the PPv/Pv conductivity
ratio increases from 1.45 at \SI{100}{\giga\pascal} to 1.55 at
\SI{135}{\giga\pascal} at \SI{300}{\kelvin}, and from 1.44 to 1.52 over
the same pressure interval at \SI{3000}{\kelvin}. The phase contrast
therefore increases moderately with pressure but changes little with
temperature because both phases exhibit a nearly inverse temperature
dependence. At \SI{135}{\giga\pascal} and \SI{2000}{\kelvin}, the present
ratio of 1.53 is close to the value of 1.58 implied by
\citet{DekuraTsuchiya2019}; classical molecular dynamics likewise predicted
an increase of about 50\% \citep{Ammann2014}. Approaches that include
higher-order anharmonic effects have generally yielded smaller contrasts:
about 25\% from phonon-quasiparticle calculations
\citep{ZhangWentzcovitch2021}, about 21\% at
\SI{140}{\giga\pascal} and \SI{4000}{\kelvin} from machine-learning
Green--Kubo simulations \citep{PengDeng2024}, and 18.2\% along a model
geotherm from deep-potential molecular dynamics \citep{Yang2022}. At room
temperature, the thermal diffusivity of PPv measured by \citet{Ohta2012} is
approximately 72\% higher than that of Pv at comparable pressure. The
present results therefore support the larger phase contrast obtained from
full-LBTE calculations and classical-potential simulations, although whether
this magnitude persists when higher-order anharmonicity and natural-mantle
chemistry are included remains uncertain. The conductivity discontinuity
along the calculated phase boundary is evaluated in
Section~\ref{sec:geotherm}.

\begin{figure}[!tbp]
  \centering
  \includegraphics[width=\columnwidth]
  {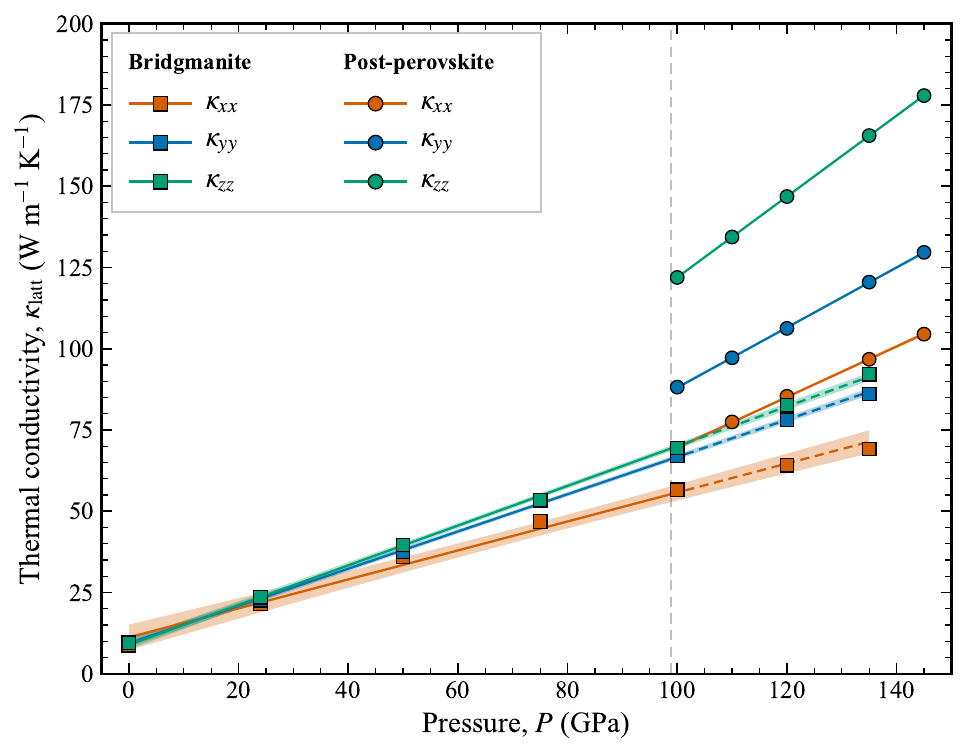}
  \caption{Rank-ordered diagonal lattice thermal-conductivity components of
  bridgmanite (squares) and post-perovskite (circles) at
  \SI{300}{\kelvin}. The plotted $xx$, $yy$, and $zz$ components denote
  $\kappa_{\min}$, $\kappa_{\mathrm{mid}}$, and $\kappa_{\max}$,
  respectively, rather than fixed Cartesian axes across the pressure
  series. For post-perovskite, these correspond to [010], [001], and [100].
  Shaded regions denote 95\% regression confidence bands. Solid lines
  denote the stable phase according to the calculated Pv--PPv boundary,
  whereas dashed lines denote its metastable continuation. The vertical
  dashed line marks the boundary extrapolated to \SI{300}{\kelvin}
  (\SI{98.8}{\giga\pascal}).}
  \label{fig:silicates_directional}
\end{figure}

The two silicates also differ markedly in their directional transport. For
Pv, the ratio $\kappa_{\max}/\kappa_{\min}$ increases from 1.09 at
\SI{24}{\giga\pascal} to 1.33 at \SI{135}{\giga\pascal} at
\SI{300}{\kelvin}, and is 1.23 at \SI{135}{\giga\pascal} and
\SI{3500}{\kelvin}. This moderate anisotropy is broadly consistent with the
approximately 25\% reported by \citet{Ghaderi2017}. PPv is considerably more
anisotropic: at every calculated state, conductivity is highest along [100],
lowest along [010], normal to the \ce{SiO6} layers, and intermediate along
[001] (Fig.~\ref{fig:silicates_directional} and Text~S6). This ordering
agrees with the calculations of \citet{DekuraTsuchiya2019},
\citet{Wang2023}, and \citet{PengDeng2024}.

The PPv ratio $\kappa_{[100]}/\kappa_{[010]}$ increases from 1.7--1.8 at
\SI{300}{\kelvin} to 1.8--2.1 between 1500 and
\SI{4000}{\kelvin}. At \SI{135}{\giga\pascal} and
\SI{3500}{\kelvin}, the [100], [010], and [001] components are approximately
1.3, 0.7, and 1.0 times the orientational average, respectively, close to
the values of \citet{DekuraTsuchiya2019} (Text~S6). Directional differences
are averaged in a randomly oriented aggregate, whereas ideal vertical
alignment of [100] or [010] in a textured PPv layer could raise or lower
radial conduction by approximately 30\%. Alignment of [001] would leave the
radial conductivity close to the orientational average.

\subsection{Phase stability and thermal conductivity along the lower-mantle geotherm}
\label{sec:geotherm}

The DFT-QHA Gibbs free energies place the Pv--PPv transition at
\SI{111.7}{\giga\pascal} at \SI{2000}{\kelvin} and
\SI{122.4}{\giga\pascal} at \SI{3500}{\kelvin}. Over this temperature
interval, the phase boundary is represented by
\begin{equation}
P_{\mathrm{tr}}(\mathrm{GPa})
=
96.62
+
0.00736\,T(\mathrm{K}),
\label{eq:pv_ppv_boundary}
\end{equation}
corresponding to a Clapeyron slope of
\SI{7.4 \pm 0.1}{\mega\pascal\per\kelvin}. This agrees, within uncertainty,
with the \SI{7.5 \pm 0.3}{\mega\pascal\per\kelvin} obtained in the DFT-QHA
prediction of \citet{TsuchiyaT2004} and is slightly lower than the PBE-QHA
value of \SI{8.1}{\mega\pascal\per\kelvin} reported by
\citet{ZhangWentzcovitch2022}. The Gibbs free-energy differences are shown in
Fig.~S7, and the boundary fit is given in Text~S7. At
\SI{2500}{\kelvin}, Eq.~\ref{eq:pv_ppv_boundary} gives a transition pressure
of \SI{115.0}{\giga\pascal}, close to the
\SI{117}{\giga\pascal} obtained with the same exchange--correlation
functional by \citet{ZhangWentzcovitch2022}. PPv is 1.3--1.5\% denser than
Pv at the same pressure and temperature, consistent with the volume
discontinuity predicted and measured for the transition
\citep{TsuchiyaT2004,Komabayashi2008}.

\begin{figure}[!tbp]
  \centering
  \includegraphics[width=\columnwidth]
  {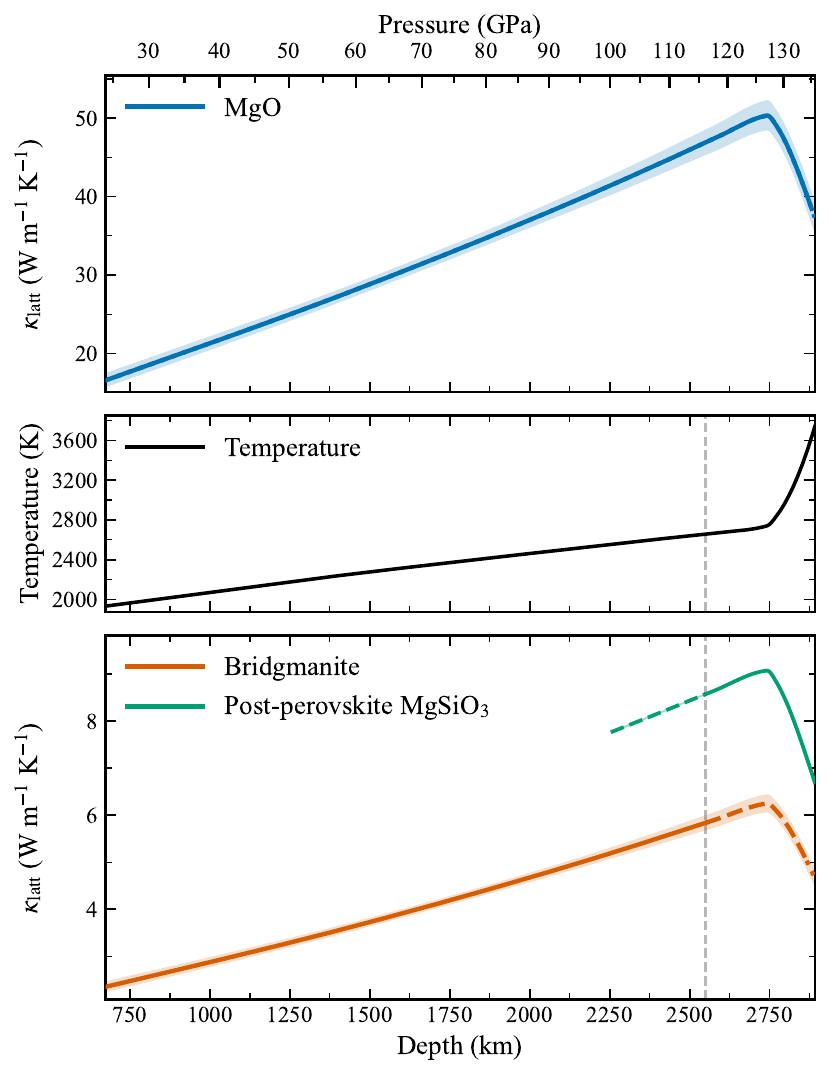}
  \caption{Lattice thermal conductivity along the adopted lower-mantle
  geotherm. The panels show the conductivity of \ce{MgO} (top), the
  geotherm temperature (middle), and the conductivities of bridgmanite and
  post-perovskite (bottom) as functions of depth; the upper axis gives
  pressure. The vertical dashed line marks the DFT-QHA phase-boundary
  intersection at \SI{116.2}{\giga\pascal}, \SI{2660}{\kelvin}, and a
  depth of approximately \SI{2550}{\kilo\metre}. Solid silicate segments
  denote the stable phase and dashed segments its metastable continuation.
  Shaded regions are 95\% confidence intervals propagated from the
  conductivity and thermal-EOS fits.}
  \label{fig:geotherm}
\end{figure}

The calculated boundary intersects the adopted Stacey--Davis geotherm at
\SI{116.2}{\giga\pascal} and \SI{2660}{\kelvin}, corresponding to a depth
of approximately \SI{2550}{\kilo\metre}, or about
\SI{340}{\kilo\metre} above the core--mantle boundary
(Fig.~\ref{fig:geotherm}). At the CMB temperature on this geotherm,
\SI{3739}{\kelvin}, the boundary, extrapolated slightly beyond the fitted
interval, lies at \SI{124.1}{\giga\pascal}, below the CMB pressure of
\SI{135.8}{\giga\pascal}. The geotherm therefore crosses the calculated
boundary only once, and PPv remains stable to the base of the mantle. This
behaviour is consistent with the results of \citet{ZhangWentzcovitch2022},
whose PBE transition temperatures at the CMB are \SI{4650}{\kelvin} within
the QHA and \SI{4310}{\kelvin} from the phonon-quasiparticle treatment, both
well above the temperature of the adopted geotherm. It differs from the
experimental boundary of \citet{Tateno2009}, whose transition temperature of
\SI{3520 \pm 70}{\kelvin} at \SI{136}{\giga\pascal} would permit a
back-transformation to Pv close to the CMB. The position and topology of the
boundary remain sensitive to the pressure scale, the treatment of
anharmonicity, and the incorporation of Fe and Al. The present QHA boundary
is therefore used to determine the stable end-member phase along the adopted
geotherm rather than as a precise prediction of the depth or thickness of
the D$^{\prime\prime}$ layer.

Along the geotherm, compression initially outweighs heating. Immediately on
the Pv side of the phase boundary, the conductivities are
\SI{46.9}{\watt\per\metre\per\kelvin} for \ce{MgO} and
\SI{5.8}{\watt\per\metre\per\kelvin} for Pv. Transformation to PPv at the
same pressure and temperature raises the silicate conductivity to
\SI{8.6}{\watt\per\metre\per\kelvin}, an increase of approximately 47\%.
With increasing depth beyond the transition, the conductivities continue to
rise, reaching maxima of \SI{50.3}{\watt\per\metre\per\kelvin} for
\ce{MgO} and \SI{9.1}{\watt\per\metre\per\kelvin} for PPv near
\SI{127}{\giga\pascal} and \SI{2740}{\kelvin}. The rapid temperature
increase across the basal thermal boundary layer then overwhelms the effect
of compression. At the CMB, the conductivities have fallen by about 26\%
from these maxima, to \SI{37.3}{\watt\per\metre\per\kelvin} for \ce{MgO}
and \SI{6.7}{\watt\per\metre\per\kelvin} for PPv. Across the profile,
\ce{MgO} remains approximately 5.5--8 times more conductive than the stable
coexisting silicate.

To place the end-member contrast in the context of a multiphase assemblage,
we calculated Hashin--Shtrikman bounds for an idealized aggregate containing
80~vol\% silicate and 20~vol\% \ce{MgO}, following the two-phase
approximation used in previous lower-mantle conductivity studies
\citep{Manthilake2011,DekuraTsuchiya2019,ZhangWentzcovitch2021}.
Immediately before the transition, the Pv-bearing aggregate has bounds of
\SIrange{8.7}{12.0}{\watt\per\metre\per\kelvin}. Replacing Pv with PPv
raises them to \SIrange{12.1}{14.5}{\watt\per\metre\per\kelvin}, an
increase of approximately 20--39\%. This increase is smaller than the 47\%
contrast between the silicate phases because the \ce{MgO} contribution is
unchanged. At the CMB, the PPv-bearing aggregate has bounds of
\SIrange{9.5}{11.4}{\watt\per\metre\per\kelvin}, compared with
\SIrange{7.0}{9.6}{\watt\per\metre\per\kelvin} for a metastable Pv-bearing
aggregate. The Pv-bearing bounds enclose the estimates of
\SI{8.5}{\watt\per\metre\per\kelvin} and
\SI{7.9}{\watt\per\metre\per\kelvin} reported by
\citet{DekuraTsuchiya2019} and \citet{ZhangWentzcovitch2021}, respectively,
whereas the PPv-bearing bounds lie between their corresponding estimates of
\SI{11.7}{\watt\per\metre\per\kelvin} and
\SI{9.4}{\watt\per\metre\per\kelvin}. The PPv-bearing bounds also encompass
the \SI{10}{\watt\per\metre\per\kelvin} used in heat-flow estimates based
on the double-crossing model \citep{Hernlund2005,Tateno2009}.

These aggregate values are intrinsic, Fe-free, lattice-only estimates and
should not be interpreted directly as the total conductivity of a pyrolitic
mantle. Machine-learning molecular-dynamics calculations indicate that
12.5~mol\% Fe lowers the lattice conductivity of Pv and PPv by approximately
10\% and 14\%, respectively \citep{Wang2023}, whereas results for Fe- and
Al-bearing bridgmanite indicate reductions of about 50\% relative to the
pure end member under lower-mantle conditions \citep{Edmund2024}. Radiative
transfer provides an additional contribution and may partly reduce the
conductivity contrast across the transition: values of approximately
\SI{2.0}{\watt\per\metre\per\kelvin} for Fe-bearing Pv and
\SI{1.1}{\watt\per\metre\per\kelvin} for Fe-bearing PPv have been reported
near the base of the mantle \citep{Lobanov2017}. The present profiles
therefore provide a controlled end-member baseline for quantifying the
effects of composition, phase proportions, radiative transport, and the
thermal structure of the lowermost mantle.

\section{Conclusions}
\label{sec:conclusions}

Phase-specific, fine-tuned MACE potentials were used in place of direct DFT
force evaluations for the finite-displacement structures in a common
workflow combining DFT-QHA equilibrium volumes, second- and third-order
force constants, and direct solution of the linearized phonon Boltzmann
transport equation for \ce{MgO}, \ce{MgSiO3} bridgmanite, and
\ce{MgSiO3} post-perovskite at lower-mantle conditions. The potentials
reproduce held-out DFT energies, forces, and stresses, quasi-harmonic volumes
to within 0.4\% at lower-mantle pressures, and compressed-state phonon
frequencies to within \SI{0.016}{\tera\hertz} (RMS). For \ce{MgO} at
\SI{0}{\giga\pascal}, using DFT instead of MACE forces in otherwise identical
transport calculations changes \kappalatt\ by 0.23\% at
\SI{300}{\kelvin} and 1.9\% at \SI{2000}{\kelvin}, less than the
finite-supercell uncertainty of the calculation itself. Within the
quasi-harmonic, three-phonon framework, the fine-tuned potentials therefore
make systematic transport calculations feasible for silicate structures
requiring of order $10^4$ displaced supercells per thermodynamic state.

All three phases show the same competition between compression and heating.
At fixed density, \kappalatt\ decreases approximately as $T^{-1}$, with
temperature exponents of 0.97--1.01, and increases with density with
exponents of 5.0--5.6. For \ce{MgO}, the full solution of the LBTE exceeds
the RTA result by up to 24\% at \SI{135}{\giga\pascal}, showing that the RTA
underestimates its conductivity in the deep mantle. When evaluated at the
experimental densities, the \ce{MgO} parameterization agrees with all nine
high-pressure, high-temperature measurements of \citet{Hasegawa2023} within
$2\sigma$, including seven within $1\sigma$. High-pressure room-temperature
measurements of \ce{MgO} and the silicates, by contrast, lie well below the
intrinsic calculations, consistent with additional scattering by
imperfections in the samples. Density-matched comparisons at high
temperature are therefore the more relevant test of intrinsic transport
calculations. The weaker temperature dependence measured for \ce{MgO},
bridgmanite, and Fe-bearing post-perovskite remains to be reconciled with
the near-$T^{-1}$ behaviour obtained here.

Post-perovskite is 44--55\% more conductive than bridgmanite at the same
pressure and temperature, and this contrast depends only weakly on
temperature. This result agrees with earlier full-LBTE calculations, whereas
simulations that sample anharmonicity directly using first-principles or
machine-learned forces give increases of approximately 18--25\%. The
magnitude of the bridgmanite--post-perovskite conductivity contrast, rather
than its sign, therefore remains a central uncertainty. Post-perovskite is
also strongly anisotropic, with
$\kappa_{[100]}>\kappa_{[001]}>\kappa_{[010]}$ and
$\kappa_{[100]}/\kappa_{[010]}=1.7$--2.1. Ideal crystallographic alignment
in a textured D$^{\prime\prime}$ layer could consequently raise or lower
radial conduction by approximately 30\% relative to the orientational
average. Bridgmanite is less anisotropic, with
$\kappa_{\max}/\kappa_{\min}\leq1.33$ over the investigated conditions.

The DFT-QHA phase boundary has a Clapeyron slope of
\SI{7.4 \pm 0.1}{\mega\pascal\per\kelvin} and intersects the adopted
geotherm once, at \SI{116.2}{\giga\pascal} and \SI{2660}{\kelvin}, about
\SI{340}{\kilo\metre} above the core--mantle boundary. The transition raises
the silicate conductivity by 47\%, to
\SI{8.6}{\watt\per\metre\per\kelvin}, whereas heating in the basal thermal
boundary layer lowers the \ce{MgO} and post-perovskite conductivities by
about 26\% from their maxima, to
\SI{37.3}{\watt\per\metre\per\kelvin} and
\SI{6.7}{\watt\per\metre\per\kelvin}, respectively, at the core--mantle
boundary. \ce{MgO} remains 5.5--8 times more conductive than the coexisting
silicate. For an aggregate of 80~vol\% silicate and 20~vol\% \ce{MgO}, the
transition raises the Hashin--Shtrikman bounds by 20--39\%, and the
post-perovskite-bearing bounds of
\SIrange{9.5}{11.4}{\watt\per\metre\per\kelvin} encompass the
\SI{10}{\watt\per\metre\per\kelvin} used in earlier heat-flow estimates.

These results describe intrinsic, three-phonon lattice transport in ideal
Fe-free end members. Iron and aluminium substitution, isotopic and defect
scattering, grain boundaries, phase proportions, crystallographic texture,
radiative transfer, and anharmonicity beyond the present treatment must all
be considered before the results are used to estimate heat flow across the
core--mantle boundary. In particular, phonon renormalization and higher-order
scattering can alter the predicted \ce{MgO} conductivity at
lowermost-mantle conditions by tens of percent. Within these limitations,
the density--temperature parameterizations in Table~S7 and the fine-tuned
potentials provide a consistent end-member reference for lower-mantle
thermal models. The resulting workflow also provides a practical route to
extending direct-LBTE calculations to chemically more complex phases, for
which systematic DFT force evaluations would be even more costly.

\section*{Code and data availability}

The fine-tuned MACE potentials developed for \ce{MgO}, \ce{MgSiO3}
bridgmanite, and \ce{MgSiO3} post-perovskite are publicly available at
\href{https://github.com/Amrt4696/mace-lattice-dynamics}
{github.com/Amrt4696/mace-lattice-dynamics}. The repository also contains
the scripts used to evaluate forces for finite-displacement structures
generated by Phono3py and to assemble the corresponding Phono3py-compatible
force files, together with step-by-step examples illustrating how the
potentials can be used within the lattice thermal-conductivity workflow
employed in this study. The calculated thermal-conductivity data and
additional data supporting the figures and analyses in this work are
available on request.

\section*{CRediT authorship contribution statement}

\textbf{Amrendra Kumar Tiwari:} Methodology, Software, Validation,
Investigation, Formal analysis, Data curation, Visualization, Writing --
original draft. \textbf{Sandro Jahn:} Conceptualization, Supervision,
Funding acquisition, Resources, Writing -- review \& editing.

\section*{Declaration of competing interest}

The authors declare that they have no known competing financial interests or
personal relationships that could have appeared to influence the work
reported in this paper.

\section*{Acknowledgements}

This work was funded by the Deutsche Forschungsgemeinschaft
(DFG, German Research Foundation) --  521315979 under
Germany's Priority Programme ``Reconstructing the deep dynamics of planet
Earth over geologic time (DeepDyn)'' SPP~2404 -- 500707704.

The authors gratefully acknowledge the computational and data resources
provided by the Leibniz Supercomputing Centre
(LRZ; \href{https://www.lrz.de}{www.lrz.de}). The authors also gratefully
acknowledge the scientific support and HPC resources provided by the
Erlangen National High Performance Computing Center (NHR@FAU) of the
Friedrich-Alexander-Universit\"at Erlangen-N\"urnberg (FAU) under the NHR
project b324dd. NHR funding is provided by federal and Bavarian state
authorities. NHR@FAU hardware is partially funded by the German Research
Foundation (DFG) -- 440719683.

\bibliographystyle{plainnat}
\bibliography{references}

@article{Ammann2014,
  author  = {Ammann, Michael W. and Walker, Andrew M. and Stackhouse, Stephen and Wookey, James and Forte, Alessandro M. and Brodholt, John P. and Dobson, David P.},
  title   = {Variation of thermal conductivity and heat flux at the {E}arth's core mantle boundary},
  journal = {Earth and Planetary Science Letters},
  year    = {2014},
  volume  = {390},
  pages   = {175--185},
  doi     = {10.1016/j.epsl.2014.01.009}
}

@inproceedings{Batatia2022,
  author    = {Batatia, Ilyes and Kov{\'a}cs, D{\'a}vid P. and Simm, Gregor N. C. and Ortner, Christoph and Cs{\'a}nyi, G{\'a}bor},
  title     = {{MACE}: Higher Order Equivariant Message Passing Neural Networks for Fast and Accurate Force Fields},
  booktitle = {Advances in Neural Information Processing Systems},
  year      = {2022},
  volume    = {35},
  pages     = {11423--11436},
  doi       = {10.52202/068431-0830}
}

@article{Batatia2025,
  author  = {Batatia, Ilyes and Benner, Philipp and Chiang, Yuan and Elena, Alin M. and Kov{\'a}cs, D{\'a}vid P. and Riebesell, Janosh and Advincula, Xavier R. and Asta, Mark and Avaylon, Matthew and Baldwin, William J. and Berger, Fabian and Bernstein, Noam and Bhowmik, Arghya and Bigi, Filippo and Blau, Samuel M. and C{\u a}rare, Vlad and Ceriotti, Michele and Chong, Sanggyu and Darby, James P. and De, Sandip and Della Pia, Flaviano and Deringer, Volker L. and Elijo{\v s}ius, Rokas and El-Machachi, Zakariya and Fako, Edvin and Falcioni, Fabio and Ferrari, Andrea C. and Gardner, John L. A. and Gawkowski, Miko{\l}aj J. and Genreith-Schriever, Annalena and George, Janine and Goodall, Rhys E. A. and Grandel, Jonas and Grey, Clare P. and Grigorev, Petr and Han, Shuang and Handley, Will and Heenen, Hendrik H. and Hermansson, Kersti and Ho, Cheuk Hin and Hofmann, Stephan and Holm, Christian and Jaafar, Jad and Jakob, Konstantin S. and Jung, Hyunwook and Kapil, Venkat and Kaplan, Aaron D. and Karimitari, Nima and Kermode, James R. and Kourtis, Panagiotis and Kroupa, Namu and Kullgren, Jolla and Kuner, Matthew C. and Kuryla, Domantas and Liepuoniute, Guoda and Lin, Chen and Margraf, Johannes T. and Magd{\u a}u, Ioan-Bogdan and Michaelides, Angelos and Moore, J. Harry and Naik, Aakash A. and Niblett, Samuel P. and Norwood, Sam Walton and O'Neill, Niamh and Ortner, Christoph and Persson, Kristin A. and Reuter, Karsten and Rosen, Andrew S. and Rosset, Louise A. M. and Schaaf, Lars L. and Schran, Christoph and Shi, Benjamin X. and Sivonxay, Eric and Stenczel, Tam{\'a}s K. and Sutton, Christopher and Svahn, Viktor and Swinburne, Thomas D. and Tilly, Jules and {van der Oord}, Cas and Vargas, Santiago and Varga-Umbrich, Eszter and Vegge, Tejs and Vondr{\'a}k, Martin and Wang, Yangshuai and Witt, William C. and Wolf, Thomas and Zills, Fabian and Cs{\'a}nyi, G{\'a}bor},
  title   = {A foundation model for atomistic materials chemistry},
  journal = {The Journal of Chemical Physics},
  year    = {2025},
  volume  = {163},
  number  = {18},
  pages   = {184110},
  doi     = {10.1063/5.0297006}
}

@article{Chabane2025,
  author  = {Chabane, Saadi and Paulatto, Lorenzo and Antonangeli, Daniele and Giura, Paola},
  title   = {Pressure- and temperature-dependent anharmonicity of {MgO}: Implications for the thermal conductivity of planetary mantles},
  journal = {Earth and Planetary Science Letters},
  year    = {2025},
  volume  = {651},
  pages   = {119170},
  doi     = {10.1016/j.epsl.2024.119170}
}

@article{Chaput2013,
  author  = {Chaput, Laurent},
  title   = {Direct solution to the linearized phonon {B}oltzmann equation},
  journal = {Physical Review Letters},
  year    = {2013},
  volume  = {110},
  number  = {26},
  pages   = {265506},
  doi     = {10.1103/PhysRevLett.110.265506}
}

@article{deKoker2010,
  author  = {{de Koker}, Nico},
  title   = {Thermal conductivity of {MgO} periclase at high pressure: Implications for the {D$''$} region},
  journal = {Earth and Planetary Science Letters},
  year    = {2010},
  volume  = {292},
  number  = {3--4},
  pages   = {392--398},
  doi     = {10.1016/j.epsl.2010.02.011}
}

@article{Dekura2013,
  author  = {Dekura, Haruhiko and Tsuchiya, Taku and Tsuchiya, Jun},
  title   = {Ab initio lattice thermal conductivity of {MgSiO$_3$} perovskite as found in {E}arth's lower mantle},
  journal = {Physical Review Letters},
  year    = {2013},
  volume  = {110},
  number  = {2},
  pages   = {025904},
  doi     = {10.1103/PhysRevLett.110.025904}
}

@article{DekuraTsuchiya2019,
  author  = {Dekura, Haruhiko and Tsuchiya, Taku},
  title   = {Lattice thermal conductivity of {MgSiO$_3$} postperovskite under the lowermost mantle conditions from ab initio anharmonic lattice dynamics},
  journal = {Geophysical Research Letters},
  year    = {2019},
  volume  = {46},
  number  = {22},
  pages   = {12919--12926},
  doi     = {10.1029/2019GL085273}
}

@article{DekuraTsuchiya2017,
  author  = {Dekura, Haruhiko and Tsuchiya, Taku},
  title   = {Ab initio lattice thermal conductivity of {MgO} from a complete solution of the linearized {B}oltzmann transport equation},
  journal = {Physical Review B},
  year    = {2017},
  volume  = {95},
  number  = {18},
  pages   = {184303},
  doi     = {10.1103/PhysRevB.95.184303}
}

@article{Edmund2024,
  author  = {Edmund, Eric and Chuvashova, Irina and Kon{\^o}pkov{\'a}, Zuzana and Husband, Rachel and Strohm, Cornelius and Appel, Karen and B{\"a}htz, Carsten and Ball, Orianna and Bouffetier, Victorien and Brugman, Kara and Buakor, Khachiwan and Chantel, Julien and Chariton, Stella and Duff, Matthew and Dwivedi, Anand and Glazyrin, Konstantin and Hosseini-Saber, S. M. A. and Jaisle, Nicolas and Laurus, Torsten and Li, Xiang and Masani, Bernhard and McHardy, James and McMahon, Malcolm and Merkel, S{\'e}bastien and Mohrbach, Katharina and Mondal, Anshuman and Morard, Guillaume and Prakapenka, Vitali B. and Prescher, Clemens and Ryu, Young-Jay and Schwinkendorf, Jan-Patrick and Tang, Minxue and Younes, Zena and Sanchez-Valle, Carmen and Liermann, Hanns-Peter and Badro, James and Lin, Jung-Fu and McWilliams, R. Stewart and Goncharov, Alexander F.},
  title   = {The thermal conductivity of bridgmanite at lower mantle conditions using a multi-technique approach},
  journal = {Journal of Geophysical Research: Solid Earth},
  year    = {2024},
  volume  = {129},
  number  = {6},
  pages   = {e2024JB028823},
  doi     = {10.1029/2024JB028823}
}

@article{Ghaderi2017,
  author  = {Ghaderi, Nahid and Zhang, Dong-Bo and Zhang, Huai and Xian, Jiawei and Wentzcovitch, Renata M. and Sun, Tao},
  title   = {Lattice thermal conductivity of {MgSiO$_3$} perovskite from first principles},
  journal = {Scientific Reports},
  year    = {2017},
  volume  = {7},
  pages   = {5417},
  doi     = {10.1038/s41598-017-05523-6}
}

@article{Guignot2007,
  author  = {Guignot, Nicolas and Andrault, Denis and Morard, Guillaume and Bolfan-Casanova, Nathalie and Mezouar, Mohamed},
  title   = {Thermoelastic properties of post-perovskite phase {MgSiO$_3$} determined experimentally at core--mantle boundary {P--T} conditions},
  journal = {Earth and Planetary Science Letters},
  year    = {2007},
  volume  = {256},
  number  = {1--2},
  pages   = {162--168},
  doi     = {10.1016/j.epsl.2007.01.025}
}

@article{Haigis2012,
  author  = {Haigis, Volker and Salanne, Mathieu and Jahn, Sandro},
  title   = {Thermal conductivity of {MgO}, {MgSiO$_3$} perovskite and post-perovskite in the {E}arth's deep mantle},
  journal = {Earth and Planetary Science Letters},
  year    = {2012},
  volume  = {355--356},
  pages   = {102--108},
  doi     = {10.1016/j.epsl.2012.09.002}
}

@article{Hanseroth2026,
  author  = {H{\"a}nseroth, Jonas and Fl{\"o}totto, Aaron and Qaisrani, Muhammad Nawaz and Dre{\ss}ler, Christian},
  title   = {Fine-Tuning Unifies Foundational Machine-Learned Interatomic Potential Architectures at ab initio Accuracy},
  journal = {The Journal of Physical Chemistry Letters},
  year    = {2026},
  volume  = {17},
  pages   = {3152--3162},
  doi     = {10.1021/acs.jpclett.5c03801}
}

@article{Hasegawa2023,
  author  = {Hasegawa, Akira and Ohta, Kenji and Yagi, Takashi and Hirose, Kei},
  title   = {Thermal conductivity of platinum and periclase under extreme conditions of pressure and temperature},
  journal = {High Pressure Research},
  year    = {2023},
  volume  = {43},
  number  = {1},
  pages   = {68--80},
  doi     = {10.1080/08957959.2023.2193892}
}

@article{Hernlund2005,
  author  = {Hernlund, John W. and Thomas, Christine and Tackley, Paul J.},
  title   = {A doubling of the post-perovskite phase boundary and structure of the {E}arth's lowermost mantle},
  journal = {Nature},
  year    = {2005},
  volume  = {434},
  number  = {7035},
  pages   = {882--886},
  doi     = {10.1038/nature03472}
}

@article{Hunt2012,
  author  = {Hunt, Simon A. and Davies, D. Rhodri and Walker, Andrew M. and McCormack, Richard J. and Wills, Andrew S. and Dobson, David P. and Li, Li},
  title   = {On the increase in thermal diffusivity caused by the perovskite to post-perovskite phase transition and its implications for mantle dynamics},
  journal = {Earth and Planetary Science Letters},
  year    = {2012},
  volume  = {319--320},
  pages   = {96--103},
  doi     = {10.1016/j.epsl.2011.12.009}
}

@article{Iitaka2004,
  author  = {Iitaka, Toshiaki and Hirose, Kei and Kawamura, Katsuyuki and Murakami, Motohiko},
  title   = {The elasticity of the {MgSiO$_3$} post-perovskite phase in the {E}arth's lowermost mantle},
  journal = {Nature},
  year    = {2004},
  volume  = {430},
  number  = {6998},
  pages   = {442--445},
  doi     = {10.1038/nature02702}
}

@article{Komabayashi2008,
  author  = {Komabayashi, Tetsuya and Hirose, Kei and Sugimura, Emiko and Sata, Nagayoshi and Ohishi, Yasuo and Dubrovinsky, Leonid S.},
  title   = {Simultaneous volume measurements of post-perovskite and perovskite in {MgSiO$_3$} and their thermal equations of state},
  journal = {Earth and Planetary Science Letters},
  year    = {2008},
  volume  = {265},
  number  = {3--4},
  pages   = {515--524},
  doi     = {10.1016/j.epsl.2007.10.036}
}

@article{KresseFurthmuller1996,
  author  = {Kresse, Georg and Furthm{\"u}ller, J{\"u}rgen},
  title   = {Efficient iterative schemes for ab initio total-energy calculations using a plane-wave basis set},
  journal = {Physical Review B},
  year    = {1996},
  volume  = {54},
  number  = {16},
  pages   = {11169--11186},
  doi     = {10.1103/PhysRevB.54.11169}
}

@article{KresseJoubert1999,
  author  = {Kresse, Georg and Joubert, Daniel},
  title   = {From ultrasoft pseudopotentials to the projector augmented-wave method},
  journal = {Physical Review B},
  year    = {1999},
  volume  = {59},
  number  = {3},
  pages   = {1758--1775},
  doi     = {10.1103/PhysRevB.59.1758}
}

@article{Loew2025,
  author  = {Loew, Antoine and Sun, Dewen and Wang, Hai-Chen and Botti, Silvana and Marques, Miguel A. L.},
  title   = {Universal machine learning interatomic potentials are ready for phonons},
  journal = {npj Computational Materials},
  year    = {2025},
  volume  = {11},
  pages   = {178},
  doi     = {10.1038/s41524-025-01650-1}
}

@article{Loew2026,
  author  = {Loew, Antoine and Schmidt, Jonathan and Botti, Silvana and Marques, Miguel A. L.},
  title   = {Universal machine learning potentials under pressure},
  journal = {Journal of Physics: Materials},
  year    = {2026},
  volume  = {9},
  number  = {1},
  pages   = {015010},
  doi     = {10.1088/2515-7639/ae2ba8}
}

@article{Manthilake2011,
  author  = {Manthilake, Geeth M. and de Koker, Nico and Frost, Daniel J. and McCammon, Catherine A.},
  title   = {Lattice thermal conductivity of lower mantle minerals and heat flux from {E}arth's core},
  journal = {Proceedings of the National Academy of Sciences of the United States of America},
  year    = {2011},
  volume  = {108},
  number  = {44},
  pages   = {17901--17904},
  doi     = {10.1073/pnas.1110594108}
}

@article{McGaughey2025,
  author  = {McGaughey, Alan J. H. and Lindsay, Lucas and Bao, Hua and Hamakawa, Tomu and Juneja, Rinkle and Li, Shouhang and Li, Wu and Masuki, Ryota and Meng, Fanchen and Meng, Han and Pandey, Tribhuwan and Shao, Cheng and Shiomi, Junichiro and Tadano, Terumasa and Togo, Atsushi and Wang, Ao and Zhang, Xinyu},
  title   = {Phonon {O}lympics: Phonon property and lattice thermal conductivity benchmarking from open-source packages},
  journal = {Journal of Applied Physics},
  year    = {2025},
  volume  = {138},
  number  = {13},
  pages   = {135108},
  doi     = {10.1063/5.0289819}
}

@article{Murakami2004,
  author  = {Murakami, Motohiko and Hirose, Kei and Kawamura, Katsuyuki and Sata, Nagayoshi and Ohishi, Yasuo},
  title   = {Post-perovskite phase transition in {MgSiO$_3$}},
  journal = {Science},
  year    = {2004},
  volume  = {304},
  number  = {5672},
  pages   = {855--858},
  doi     = {10.1126/science.1095932}
}

@article{OganovOno2004,
  author  = {Oganov, Artem R. and Ono, Shigeaki},
  title   = {Theoretical and experimental evidence for a post-perovskite phase of {MgSiO$_3$} in {E}arth's {D$''$} layer},
  journal = {Nature},
  year    = {2004},
  volume  = {430},
  number  = {6998},
  pages   = {445--448},
  doi     = {10.1038/nature02701}
}

@article{Ohta2012,
  author  = {Ohta, Kenji and Yagi, Takashi and Taketoshi, Naoyuki and Hirose, Kei and Komabayashi, Tetsuya and Baba, Tetsuya and Ohishi, Yasuo and Hernlund, John},
  title   = {Lattice thermal conductivity of {MgSiO$_3$} perovskite and post-perovskite at the core--mantle boundary},
  journal = {Earth and Planetary Science Letters},
  year    = {2012},
  volume  = {349--350},
  pages   = {109--115},
  doi     = {10.1016/j.epsl.2012.06.043}
}

@article{Okuda2020,
  author  = {Okuda, Yoshiyuki and Ohta, Kenji and Hasegawa, Akira and Yagi, Takashi and Hirose, Kei and Kawaguchi, Saori I. and Ohishi, Yasuo},
  title   = {Thermal conductivity of {Fe}-bearing post-perovskite in the {E}arth's lowermost mantle},
  journal = {Earth and Planetary Science Letters},
  year    = {2020},
  volume  = {547},
  pages   = {116466},
  doi     = {10.1016/j.epsl.2020.116466}
}

@article{PengDeng2024,
  author  = {Peng, Yihang and Deng, Jie},
  title   = {Thermal conductivity of {MgSiO$_3$--H$_2$O} system determined by machine learning potentials},
  journal = {Geophysical Research Letters},
  year    = {2024},
  volume  = {51},
  number  = {5},
  pages   = {e2023GL107245},
  doi     = {10.1029/2023GL107245}
}

@article{Perdew1996,
  author  = {Perdew, John P. and Burke, Kieron and Ernzerhof, Matthias},
  title   = {Generalized Gradient Approximation Made Simple},
  journal = {Physical Review Letters},
  year    = {1996},
  volume  = {77},
  number  = {18},
  pages   = {3865--3868},
  doi     = {10.1103/PhysRevLett.77.3865}
}

@article{Pota2026,
  author  = {P{\'o}ta, Bal{\'a}zs and Ahlawat, Paramvir and Cs{\'a}nyi, G{\'a}bor and Simoncelli, Michele},
  title   = {Thermal conductivity predictions with foundation atomistic models},
  journal = {Nature Communications},
  year    = {2026},
  volume  = {17},
  number  = {1},
  pages   = {10566},
  doi     = {10.1038/s41467-026-76391-w}
}

@book{StaceyDavis2008,
  author    = {Stacey, Frank D. and Davis, Paul M.},
  title     = {Physics of the Earth},
  edition   = {Fourth},
  publisher = {Cambridge University Press},
  address   = {Cambridge},
  year      = {2008},
  doi       = {10.1017/CBO9780511812910}
}

@article{Stackhouse2010,
  author  = {Stackhouse, Stephen and Stixrude, Lars and Karki, Bijaya B.},
  title   = {Thermal conductivity of periclase ({MgO}) from first principles},
  journal = {Physical Review Letters},
  year    = {2010},
  volume  = {104},
  number  = {20},
  pages   = {208501},
  doi     = {10.1103/PhysRevLett.104.208501}
}

@article{Tamura1983,
  author  = {Tamura, Shin-ichiro},
  title   = {Isotope scattering of dispersive phonons in {Ge}},
  journal = {Physical Review B},
  year    = {1983},
  volume  = {27},
  number  = {2},
  pages   = {858--866},
  doi     = {10.1103/PhysRevB.27.858}
}

@article{Tang2014,
  author  = {Tang, Xiaoli and Ntam, Moses C. and Dong, Jianjun and Rainey, Emma S. G. and Kavner, Abby},
  title   = {The thermal conductivity of {E}arth's lower mantle},
  journal = {Geophysical Research Letters},
  year    = {2014},
  volume  = {41},
  number  = {8},
  pages   = {2746--2752},
  doi     = {10.1002/2014GL059385}
}

@article{TangDong2010,
  author  = {Tang, Xiaoli and Dong, Jianjun},
  title   = {Lattice thermal conductivity of {MgO} at conditions of {E}arth's interior},
  journal = {Proceedings of the National Academy of Sciences of the United States of America},
  year    = {2010},
  volume  = {107},
  number  = {10},
  pages   = {4539--4543},
  doi     = {10.1073/pnas.0907194107}
}

@article{Tateno2009,
  author  = {Tateno, Shigehiko and Hirose, Kei and Sata, Nagayoshi and Ohishi, Yasuo},
  title   = {Determination of post-perovskite phase transition boundary up to 4400 {K} and implications for thermal structure in {D$''$} layer},
  journal = {Earth and Planetary Science Letters},
  year    = {2009},
  volume  = {277},
  number  = {1--2},
  pages   = {130--136},
  doi     = {10.1016/j.epsl.2008.10.004}
}

@article{Togo2015,
  author  = {Togo, Atsushi and Chaput, Laurent and Tanaka, Isao},
  title   = {Distributions of phonon lifetimes in {B}rillouin zones},
  journal = {Physical Review B},
  year    = {2015},
  volume  = {91},
  number  = {9},
  pages   = {094306},
  doi     = {10.1103/PhysRevB.91.094306}
}

@article{Togo2023,
  author  = {Togo, Atsushi and Chaput, Laurent and Tadano, Terumasa and Tanaka, Isao},
  title   = {Implementation strategies in phonopy and phono3py},
  journal = {Journal of Physics: Condensed Matter},
  year    = {2023},
  volume  = {35},
  number  = {35},
  pages   = {353001},
  doi     = {10.1088/1361-648X/acd831}
}

@article{TsuchiyaJ2005,
  author  = {Tsuchiya, Jun and Tsuchiya, Taku and Wentzcovitch, Renata M.},
  title   = {Vibrational and thermodynamic properties of {MgSiO$_3$} postperovskite},
  journal = {Journal of Geophysical Research: Solid Earth},
  year    = {2005},
  volume  = {110},
  number  = {B2},
  pages   = {B02204},
  doi     = {10.1029/2004JB003409}
}

@article{TsuchiyaT2004,
  author  = {Tsuchiya, Taku and Tsuchiya, Jun and Umemoto, Koichiro and Wentzcovitch, Renata M.},
  title   = {Phase transition in {MgSiO$_3$} perovskite in the {E}arth's lower mantle},
  journal = {Earth and Planetary Science Letters},
  year    = {2004},
  volume  = {224},
  number  = {3--4},
  pages   = {241--248},
  doi     = {10.1016/j.epsl.2004.05.017}
}

@article{Wang2023,
  author  = {Wang, Dong and Wu, Zhongqing and Deng, Xin},
  title   = {Thermal conductivity of {Fe}-bearing bridgmanite and post-perovskite: Implications for the heat flux from the core},
  journal = {Earth and Planetary Science Letters},
  year    = {2023},
  volume  = {621},
  pages   = {118368},
  doi     = {10.1016/j.epsl.2023.118368}
}

@article{Yang2022,
  author  = {Yang, Fenghu and Zeng, Qiyu and Chen, Bo and Kang, Dongdong and Zhang, Shen and Wu, Jianhua and Yu, Xiaoxiang and Dai, Jiayu},
  title   = {Lattice thermal conductivity of {MgSiO$_3$} perovskite and post-perovskite under lower mantle conditions calculated by deep potential molecular dynamics},
  journal = {Chinese Physics Letters},
  year    = {2022},
  volume  = {39},
  number  = {11},
  pages   = {116301},
  doi     = {10.1088/0256-307X/39/11/116301}
}

@article{ZhangWentzcovitch2021,
  author  = {Zhang, Zhen and Wentzcovitch, Renata M.},
  title   = {Ab initio lattice thermal conductivity of {MgSiO$_3$} across the perovskite--postperovskite phase transition},
  journal = {Physical Review B},
  year    = {2021},
  volume  = {103},
  number  = {14},
  pages   = {144103},
  doi     = {10.1103/PhysRevB.103.144103}
}

@article{ZhangWentzcovitch2022,
  author  = {Zhang, Zhen and Wentzcovitch, Renata M.},
  title   = {Anharmonic thermodynamic properties and phase boundary across the postperovskite transition in {MgSiO$_3$}},
  journal = {Physical Review B},
  year    = {2022},
  volume  = {106},
  number  = {5},
  pages   = {054103},
  doi     = {10.1103/PhysRevB.106.054103}
}

@article{Deng2023,
  author  = {Deng, Bowen and Zhong, Peichen and Jun, KyuJung and Riebesell, Janosh and Han, Kevin and Bartel, Christopher J. and Ceder, Gerbrand},
  title   = {{CHGNet} as a pretrained universal neural network potential for charge-informed atomistic modelling},
  journal = {Nature Machine Intelligence},
  year    = {2023},
  volume  = {5},
  pages   = {1031--1041},
  doi     = {10.1038/s42256-023-00716-3}
}

@article{BarrosLuque2026,
  author  = {Barros-Luque, Luis and Shuaibi, Muhammed and Fu, Xiang and Wood, Brandon M. and Dzamba, Misko and Gao, Meng and Rizvi, Ammar and Uyttendaele, Matt and Zitnick, C. Lawrence and Ulissi, Zachary W.},
  title   = {The {Open Materials 2024} ({OMat24}) inorganic materials dataset and models},
  journal = {Nature Computational Science},
  year    = {2026},
  volume  = {6},
  pages   = {642--652},
  doi     = {10.1038/s43588-026-00996-w}
}

@article{Birch1947,
  author  = {Birch, Francis},
  title   = {Finite Elastic Strain of Cubic Crystals},
  journal = {Physical Review},
  year    = {1947},
  volume  = {71},
  number  = {11},
  pages   = {809--824},
  doi     = {10.1103/PhysRev.71.809}
}

@article{Dewaele2000,
  author  = {Dewaele, Agn{\`e}s and Fiquet, Guillaume and Andrault, Denis and H{\"a}usermann, Daniel},
  title   = {{P--V--T} equation of state of periclase from synchrotron radiation measurements},
  journal = {Journal of Geophysical Research: Solid Earth},
  year    = {2000},
  volume  = {105},
  number  = {B2},
  pages   = {2869--2877},
  doi     = {10.1029/1999JB900364}
}

@article{StixrudeLithgowBertelloni2005,
  author  = {Stixrude, Lars and Lithgow-Bertelloni, Carolina},
  title   = {Thermodynamics of mantle minerals---{I}. Physical properties},
  journal = {Geophysical Journal International},
  year    = {2005},
  volume  = {162},
  number  = {2},
  pages   = {610--632},
  doi     = {10.1111/j.1365-246X.2005.02642.x}
}

@article{Tange2009,
  author  = {Tange, Yoshinori and Nishihara, Yu and Tsuchiya, Taku},
  title   = {Unified analyses for {P--V--T} equation of state of {MgO}: A solution for pressure-scale problems in high {P--T} experiments},
  journal = {Journal of Geophysical Research: Solid Earth},
  year    = {2009},
  volume  = {114},
  number  = {B3},
  pages   = {B03208},
  doi     = {10.1029/2008JB005813}
}

@incollection{Utsumi1998,
  author    = {Utsumi, Wataru and Weidner, Donald J. and Liebermann, Robert C.},
  title     = {Volume Measurement of {MgO} at High Pressures and High Temperatures},
  booktitle = {Properties of Earth and Planetary Materials at High Pressure and Temperature},
  editor    = {Manghnani, Murli H. and Yagi, Takehiko},
  series    = {Geophysical Monograph Series},
  volume    = {101},
  publisher = {American Geophysical Union},
  address   = {Washington, D.C.},
  year      = {1998},
  pages     = {327--333},
  doi       = {10.1029/GM101p0327}
}

@article{Criniti2024,
  author  = {Criniti, Giacomo and Boffa Ballaran, Tiziana and Kurnosov, Alexander and Liu, Zhaodong and Glazyrin, Konstantin and Merlini, Marco and Hanfland, Michael and Frost, Daniel J.},
  title   = {Thermal Equation of State and Structural Evolution of {Al}-Bearing Bridgmanite},
  journal = {Journal of Geophysical Research: Solid Earth},
  year    = {2024},
  volume  = {129},
  number  = {1},
  pages   = {e2023JB026879},
  doi     = {10.1029/2023JB026879}
}

@article{Kudoh1987,
  author  = {Kudoh, Y. and Ito, E. and Takeda, H.},
  title   = {Effect of Pressure on the Crystal Structure of Perovskite-Type {MgSiO$_3$}},
  journal = {Physics and Chemistry of Minerals},
  year    = {1987},
  volume  = {14},
  number  = {4},
  pages   = {350--354},
  doi     = {10.1007/BF00309809}
}

@article{Utsumi1995,
  author  = {Utsumi, Wataru and Funamori, Nobumasa and Yagi, Takehiko and Ito, Eiji and Kikegawa, Takumi and Shimomura, Osamu},
  title   = {Thermal Expansivity of {MgSiO$_3$} Perovskite Under High Pressures up to 20 {GPa}},
  journal = {Geophysical Research Letters},
  year    = {1995},
  volume  = {22},
  number  = {9},
  pages   = {1005--1008},
  doi     = {10.1029/95GL00584}
}

@article{Fiquet2000,
  author  = {Fiquet, Guillaume and Dewaele, Agn{\`e}s and Andrault, Denis and Kunz, Martin and Le Bihan, Thierry},
  title   = {Thermoelastic Properties and Crystal Structure of {MgSiO$_3$} Perovskite at Lower Mantle Pressure and Temperature Conditions},
  journal = {Geophysical Research Letters},
  year    = {2000},
  volume  = {27},
  number  = {1},
  pages   = {21--24},
  doi     = {10.1029/1999GL008397}
}

@article{Dalton2013,
  author  = {Dalton, Douglas Allen and Hsieh, Wen-Pin and Hohensee, Gregory T. and Cahill, David G. and Goncharov, Alexander F.},
  title   = {Effect of Mass Disorder on the Lattice Thermal Conductivity of {MgO} Periclase under Pressure},
  journal = {Scientific Reports},
  year    = {2013},
  volume  = {3},
  pages   = {2400},
  doi     = {10.1038/srep02400}
}

@article{Hofmeister2014,
  author  = {Hofmeister, Anne M.},
  title   = {Thermal Diffusivity and Thermal Conductivity of Single-Crystal {MgO} and {Al$_2$O$_3$} and Related Compounds as a Function of Temperature},
  journal = {Physics and Chemistry of Minerals},
  year    = {2014},
  volume  = {41},
  pages   = {361--371},
  doi     = {10.1007/s00269-014-0655-3}
}

@article{Imada2014,
  author  = {Imada, Saori and Ohta, Kenji and Yagi, Takashi and Hirose, Kei and Yoshida, Hideto and Nagahara, Hiroko},
  title   = {Measurements of Lattice Thermal Conductivity of {MgO} to Core-Mantle Boundary Pressures},
  journal = {Geophysical Research Letters},
  year    = {2014},
  volume  = {41},
  number  = {13},
  pages   = {4542--4547},
  doi     = {10.1002/2014GL060423}
}

@article{Kanamori1968,
  author  = {Kanamori, Hiroo and Fujii, Naoyuki and Mizutani, Hitoshi},
  title   = {Thermal Diffusivity Measurement of Rock-Forming Minerals from 300$^{\circ}$ to 1100$^{\circ}${K}},
  journal = {Journal of Geophysical Research},
  year    = {1968},
  volume  = {73},
  number  = {2},
  pages   = {595--605},
  doi     = {10.1029/JB073i002p00595}
}

@article{Speziale2001,
  author  = {Speziale, Sergio and Zha, Chang-Sheng and Duffy, Thomas S. and Hemley, Russell J. and Mao, Ho-kwang},
  title   = {Quasi-Hydrostatic Compression of Magnesium Oxide to 52 {GPa}: Implications for the Pressure-Volume-Temperature Equation of State},
  journal = {Journal of Geophysical Research: Solid Earth},
  year    = {2001},
  volume  = {106},
  number  = {B1},
  pages   = {515--528},
  doi     = {10.1029/2000JB900318}
}

@book{Touloukian1973,
  author    = {Touloukian, Y. S. and Powell, R. W. and Ho, C. Y. and Nicolaou, M. C.},
  title     = {Thermal Diffusivity},
  series    = {Thermophysical Properties of Matter: The {TPRC} Data Series},
  volume    = {10},
  publisher = {IFI/Plenum},
  address   = {New York},
  year      = {1973}
}

@article{Osako1991,
  author  = {Osako, Masahiro and Ito, Eiji},
  title   = {Thermal Diffusivity of {MgSiO$_3$} Perovskite},
  journal = {Geophysical Research Letters},
  year    = {1991},
  volume  = {18},
  number  = {2},
  pages   = {239--242},
  doi     = {10.1029/91GL00212}
}

@article{Stackhouse2015,
  author  = {Stackhouse, Stephen and Stixrude, Lars and Karki, Bijaya B.},
  title   = {First-Principles Calculations of the Lattice Thermal Conductivity of the Lower Mantle},
  journal = {Earth and Planetary Science Letters},
  year    = {2015},
  volume  = {427},
  pages   = {11--17},
  doi     = {10.1016/j.epsl.2015.06.050}
}

@article{Lobanov2017,
  author  = {Lobanov, Sergey S. and Holtgrewe, Nicholas and Lin, Jung-Fu and Goncharov, Alexander F.},
  title   = {Radiative Conductivity and Abundance of Post-Perovskite in the Lowermost Mantle},
  journal = {Earth and Planetary Science Letters},
  year    = {2017},
  volume  = {479},
  pages   = {43--49},
  doi     = {10.1016/j.epsl.2017.09.016}
}

@article{HashinShtrikman1962,
  author  = {Hashin, Zvi and Shtrikman, Shmuel},
  title   = {A Variational Approach to the Theory of the Effective Magnetic Permeability of Multiphase Materials},
  journal = {Journal of Applied Physics},
  year    = {1962},
  volume  = {33},
  number  = {10},
  pages   = {3125--3131},
  doi     = {10.1063/1.1728579}
}

@article{Sangster1970,
  author  = {Sangster, M. J. L. and Peckham, G. and Saunderson, D. H.},
  title   = {Lattice Dynamics of Magnesium Oxide},
  journal = {Journal of Physics C: Solid State Physics},
  year    = {1970},
  volume  = {3},
  number  = {5},
  pages   = {1026--1036},
  doi     = {10.1088/0022-3719/3/5/017}
}

@article{Zhang2017,
  author  = {Zhang, Dong-Bo and Allen, Philip B. and Sun, Tao and Wentzcovitch, Renata M.},
  title   = {Thermal Conductivity from Phonon Quasiparticles with Subminimal Mean Free Path in {MgSiO$_3$} Perovskite},
  journal = {Physical Review B},
  year    = {2017},
  volume  = {96},
  number  = {10},
  pages   = {100302},
  doi     = {10.1103/PhysRevB.96.100302}
}

@article{Wehinger2016,
  author  = {Wehinger, Bj{\"o}rn and Bosak, Alexe{\"i} and
             Nazzareni, Sabrina and Antonangeli, Daniele and
             Mirone, Alessandro and Chaplot, Samrath Lal and
             Mittal, Ranjan and Ohtani, Eiji and Shatskiy, Anton and
             Saxena, Surendra and Ghose, Subrata and Krisch, Michael},
  title   = {Dynamical and elastic properties of {MgSiO$_3$} perovskite
             (bridgmanite)},
  journal = {Geophysical Research Letters},
  year    = {2016},
  volume  = {43},
  pages   = {2568--2575},
  doi     = {10.1002/2016GL067970}
}

@article{Bartok2013SOAP,
  author  = {Bart{\'o}k, Albert P. and Kondor, Risi and Cs{\'a}nyi, G{\'a}bor},
  title   = {On representing chemical environments},
  journal = {Physical Review B},
  year    = {2013},
  volume  = {87},
  number  = {18},
  pages   = {184115},
  doi     = {10.1103/PhysRevB.87.184115}
}

@article{Imbalzano2018,
  author  = {Imbalzano, Giulio and Anelli, Andrea and Giofr{\'e}, Daniele and
             Klees, Sinja and Behler, J{\"o}rg and Ceriotti, Michele},
  title   = {Automatic selection of atomic fingerprints and reference
             configurations for machine-learning potentials},
  journal = {The Journal of Chemical Physics},
  year    = {2018},
  volume  = {148},
  number  = {24},
  pages   = {241730},
  doi     = {10.1063/1.5024611}
}

@article{Himanen2020,
  author  = {Himanen, Lauri and J{\"a}ger, Marc O. J. and
             Morooka, Eiaki V. and Federici Canova, Filippo and
             Ranawat, Yashasvi S. and Gao, David Z. and Rinke, Patrick and
             Foster, Adam S.},
  title   = {{DScribe}: Library of descriptors for machine learning in
             materials science},
  journal = {Computer Physics Communications},
  year    = {2020},
  volume  = {247},
  pages   = {106949},
  doi     = {10.1016/j.cpc.2019.106949}
}

@article{Qiu2025,
  author  = {Qiu, Rong and Zeng, Qiyu and Han, Jinsen and Chen, Ke and
             Kang, Dongdong and Yu, Xiaoxiang and Dai, Jiayu},
  title   = {Coupled temperature-density dependence of lattice thermal
             conductivity of {MgO} at extreme conditions},
  journal = {Physical Review B},
  year    = {2025},
  volume  = {111},
  number  = {6},
  pages   = {064103},
  doi     = {10.1103/PhysRevB.111.064103}
}

@article{Han2023MgO,
  author  = {Han, Zherui and Xiong, Zixin and Riffe, William T. and
             Schonfeld, Hunter B. and Segovia, Mauricio and Song, Jiawei and
             Wang, Haiyan and Xu, Xianfan and Hopkins, Patrick E. and
             Marconnet, Amy and Ruan, Xiulin},
  title   = {Predictions and measurements of thermal conductivity of ceramic
             materials at high temperature},
  journal = {Physical Review B},
  year    = {2023},
  volume  = {108},
  number  = {18},
  pages   = {184306},
  doi     = {10.1103/PhysRevB.108.184306}
}

\end{document}


\maketitle

\section*{Contents of this Supporting Information}

\begin{itemize}
  \item Text S1: DFT reference databases and MACE fine-tuning
  \item Text S2: Quasi-harmonic calculations and pressure--volume relations
  \item Text S3: LBTE formalism and numerical convergence
  \item Text S4: Conductivity parameterizations and geotherm construction
  \item Text S5: Additional validation of the fine-tuned potentials
  \item Text S6: Directional conductivity of bridgmanite and post-perovskite
  \item Text S7: Bridgmanite--post-perovskite phase stability
  \item Tables S1--S8
  \item Figures S1--S7
\end{itemize}

\clearpage

\section*{Text S1. DFT reference databases and MACE fine-tuning}

The three reference databases were constructed using the same general
strategy. AIMD trajectories sampled the pressure--temperature range relevant
to each phase. Because consecutive AIMD frames are strongly correlated,
candidate snapshots were represented by species-resolved, configuration-level
smooth overlap of atomic positions (SOAP) descriptors. These descriptors were
constructed by averaging the atomic SOAP vectors separately for each chemical
species and concatenating the resulting species-level vectors. The SOAP descriptors were computed using DScribe \citep{Himanen2020} with a cutoff radius of
\SI{5.0}{\angstromunit}, $n_{\max}=6$, $l_{\max}=4$, and a Gaussian width of
\SI{0.5}{\angstromunit}. Farthest-point sampling was then applied independently
to each trajectory in this descriptor space to select structurally diverse
snapshots while discarding redundant configurations
\citep{Bartok2013SOAP,Imbalzano2018}. Static configurations additionally
sampled homogeneous strains, systematic volume changes, random atomic
distortions, and finite displacements associated with the harmonic and cubic
force constants. Each selected configuration was labelled with its DFT
energy, atomic forces, and stress tensor. Calculation-specific settings are
collected in Table~\ref{tab:s1_settings}.

\begin{table}[!htbp]
\centering
\small
\caption{Calculation-specific DFT, AIMD, linear-response, and MACE
fine-tuning settings. The Pv and PPv calculations use the same PAW datasets
for Mg, Si, and O.}
\label{tab:s1_settings}
\begin{tabular}{>{\raggedright\arraybackslash}p{0.30\textwidth}
                >{\centering\arraybackslash}p{0.18\textwidth}
                >{\centering\arraybackslash}p{0.20\textwidth}
                >{\centering\arraybackslash}p{0.20\textwidth}}
\toprule
Setting & \ce{MgO} & Bridgmanite & Post-perovskite \\
\midrule
PAW datasets &
Mgpv (13Apr2007), O (08Apr2002) &
Mgpv (13Apr2007), Si (05Jan2001), O (08Apr2002) &
Same as Pv \\
Reference-database plane-wave cutoff (eV) & 600 & 650 & 650 \\
AIMD cell size (atoms) & 64 & 160 & 160 \\
AIMD $k$-point sampling &
$2\times2\times2$ & $\Gamma$ point & $\Gamma$ point \\
AIMD electronic convergence (eV) &
$10^{-6}$ & $10^{-6}$ & $10^{-6}$ \\
AIMD time step (fs) & 0.5 & 1.0 & 1.0 \\
Born-charge cell (atoms) & 2 & 20 & 20 \\
Born-charge $k$-point mesh &
$8\times8\times8$ &
$6\times4\times6$ &
$10\times3\times4$ \\
Foundation model &
MACE-MPA-0 medium & MACE-MPA-0 medium & MACE-MPA-0 medium \\
Numerical precision & float64 & float64 & float64 \\
Optimizer & Adam (AMSGrad) & Adam (AMSGrad) & Adam (AMSGrad) \\
Initial learning rate &
$5\times10^{-5}$ & $5\times10^{-5}$ & $5\times10^{-5}$ \\
Batch size & 2 per GPU & 2 per GPU & 2 per GPU \\
GPUs used for fine-tuning & 4 & 2 & 4 \\
Energy:force:stress weights &
$1{:}10{:}100$ & $1{:}10{:}100$ & $1{:}10{:}100$ \\
Maximum epochs & 1000 & 1000 & 1000 \\
\bottomrule
\end{tabular}
\end{table}

Configurations were partitioned by trajectory or structural series rather
than individually. Consequently, snapshots from the same AIMD trajectory and
members of the same strain, volume, or displacement series were kept within a
single subset. This grouped split reduces information leakage between the
training, validation, and test sets. The resulting database sizes and
thermodynamic coverage are given in Table~\ref{tab:s2_databases}.

\begin{table}[!htbp]
\centering
\small
\caption{Size, thermodynamic coverage, and grouped partitioning of the DFT
reference databases.}
\label{tab:s2_databases}
\begin{tabular}{lccc}
\toprule
Dataset & \ce{MgO} & Bridgmanite & Post-perovskite \\
\midrule
Total configurations & 3528 & 4612 & 3575 \\
Training & 2822 & 3758 & 2879 \\
Validation & 353 & 470 & 348 \\
Test & 353 & 384 & 348 \\
Pressure range (GPa) & 0--135 & 0--135 & 100--145 \\
Temperature range (K) & 300--3500 & 300--3500 & 300--4000 \\
\bottomrule
\end{tabular}
\end{table}

\begin{figure}[!htbp]
  \centering
  \includegraphics[width=\textwidth]{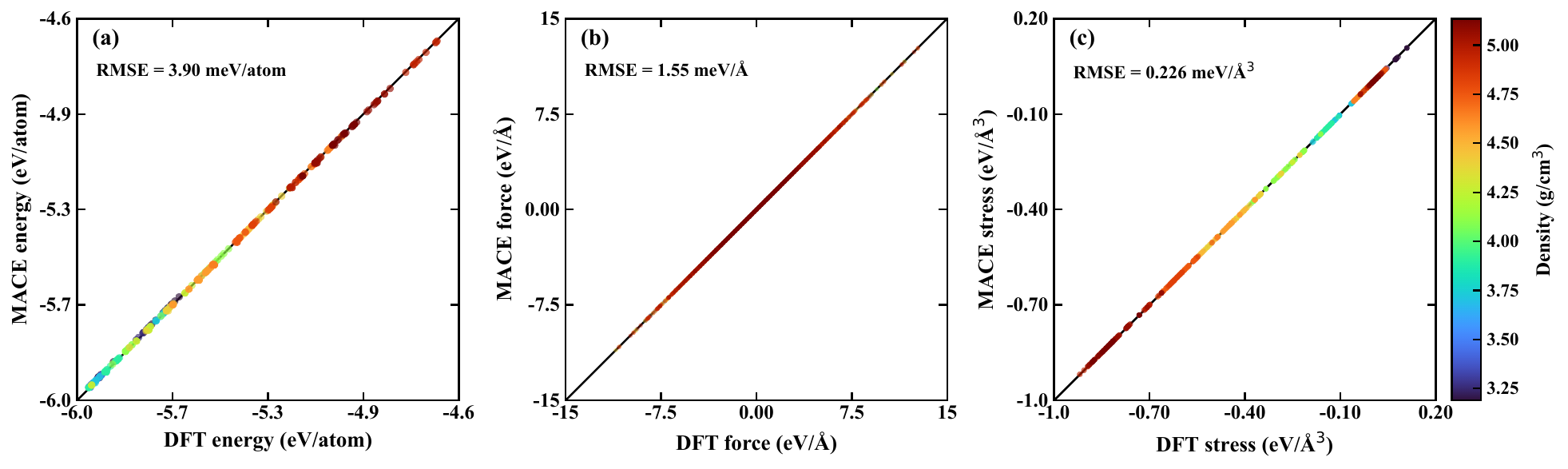}
  \caption{DFT--MACE parity plots for the independent \ce{MgO} test set:
  (a) energies, (b) atomic-force components, and (c) stress components.
  Diagonal lines mark perfect agreement, and points are coloured by density.}
  \label{fig:s1_mgo_parity}
\end{figure}

\begin{figure}[!htbp]
  \centering
  \includegraphics[width=\textwidth]{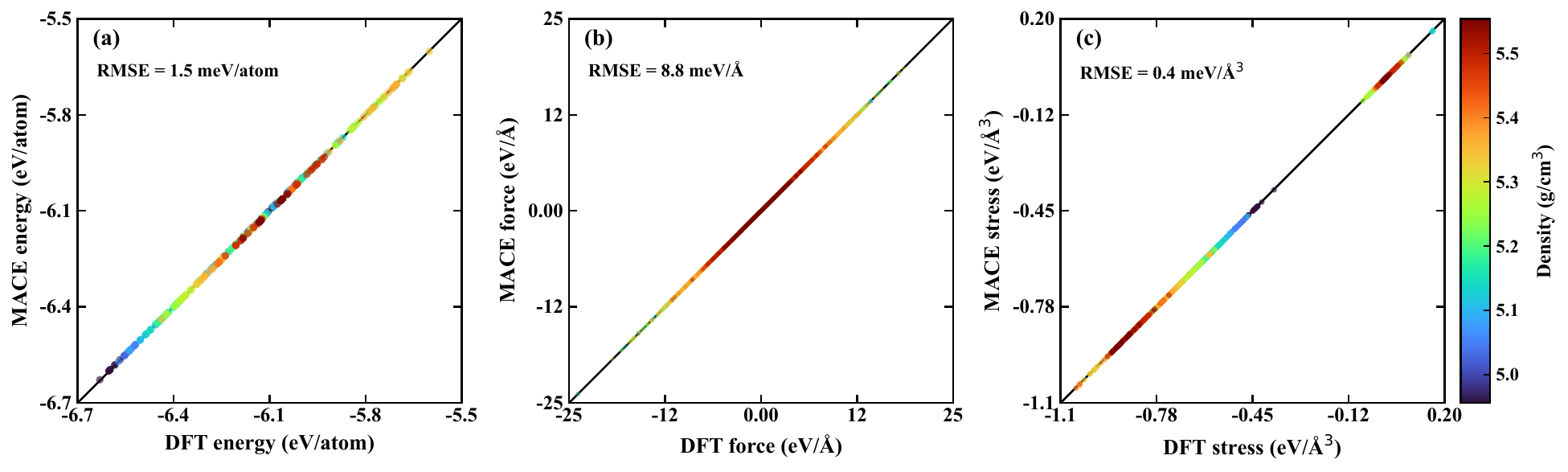}
  \caption{DFT--MACE parity plots for the independent post-perovskite test
  set: (a) energies, (b) atomic-force components, and (c) stress components.
  Diagonal lines mark perfect agreement, and points are coloured by density.}
  \label{fig:s2_ppv_parity}
\end{figure}

For each phase, the MACE-MPA-0 medium foundation model was fine-tuned using
MACE~0.3.16 in double precision \citep{Batatia2022,Batatia2025}. The same
architecture and optimization protocol were used throughout. Training used
the AMSGrad variant of Adam, an initial learning rate of $5\times10^{-5}$, an
exponential moving average with decay 0.99999, and relative
energy--force--stress weights of $1{:}10{:}100$. The held-out test subsets
were not used for optimization or model selection.
Parity plots for the independent MgO and PPv test sets are shown in
Figs.~\ref{fig:s1_mgo_parity} and~\ref{fig:s2_ppv_parity}, respectively;
the corresponding Pv comparison is shown in Fig.~1 of the main text.

\begin{table}[!htbp]
\centering
\small
\caption{Performance of the production MACE potentials on the held-out DFT
test sets. The RMS magnitudes of the DFT force components are
\SI{0.867}{\electronvolt\per\angstromunit},
\SI{1.39}{\electronvolt\per\angstromunit}, and
\SI{2.04}{\electronvolt\per\angstromunit} for \ce{MgO}, bridgmanite, and
post-perovskite, respectively; these values define the denominators of the
relative force errors reported in the main text.}
\label{tab:s3_test}
\begin{tabular}{lccc}
\toprule
Quantity & \ce{MgO} & Bridgmanite & Post-perovskite \\
\midrule
Energy RMSE (meV atom$^{-1}$)   & 3.90  & 5.84  & 1.54  \\
Energy MAE (meV atom$^{-1}$)    & 3.19  & 5.17  & 1.27  \\
Force RMSE (meV \AA$^{-1}$)     & 1.55  & 7.71  & 8.78  \\
Force MAE (meV \AA$^{-1}$)      & 0.611 & 3.86  & 4.57  \\
Stress RMSE (meV \AA$^{-3}$)    & 0.226 & 0.302 & 0.393 \\
Stress MAE (meV \AA$^{-3}$)     & 0.147 & 0.166 & 0.227 \\
\bottomrule
\end{tabular}
\end{table}

\section*{Text S2. Quasi-harmonic calculations and pressure--volume relations}

Pressure- and temperature-dependent equilibrium volumes were obtained by
minimizing the quasi-harmonic Gibbs free energy defined in the main text.
The vibrational free energy includes the zero-point and finite-temperature
harmonic contributions; finite-temperature electronic contributions were
neglected.

Harmonic QHA calculations employed a $4\times4\times4$ supercell of the
two-atom \ce{MgO} primitive cell and $2\times2\times2$ supercells of the
20-atom Pv and PPv cells. These supercells contain 128, 160, and 160 atoms,
respectively. A displacement amplitude of \SI{0.01}{\angstromunit} and a
$30\times30\times30$ phonon $q$-point mesh were used for all three phases.
Electronic Brillouin-zone sampling was converged separately for each phase
and held fixed across its volume series. For PPv, the 20-atom static cells
and 160-atom displaced supercells were sampled using $8\times4\times4$ and
$4\times2\times2$ Monkhorst--Pack meshes, respectively. For Pv and PPv,
internal atomic positions were relaxed at fixed cell before the harmonic
displaced structures were generated.

Independent QHA free-energy calculations for \ce{MgO} and Pv were performed
at 0, 24, 50, 75, 100, and \SI{135}{\giga\pascal}. For PPv, calculations
were performed at 100, 110, 120, 135, and \SI{145}{\giga\pascal}. At each
pressure, the sampled volume range bracketed the free-energy minimum over the
required temperature interval. Equilibrium volumes at intermediate pressures
used in the transport calculations were obtained from the fitted thermal
equations of state (EOS).

For \ce{MgO} and Pv, the \SI{298}{\kelvin} compression curves were fitted
with a third-order Birch--Murnaghan equation of state \citep{Birch1947}.
The resulting reference-state parameters were held fixed while the thermal
parameters were optimized using the finite-strain Debye formulation of
\citet{StixrudeLithgowBertelloni2005}. Only the independently calculated QHA
pressure curves entered the optimization; additional intermediate-pressure
curves were used for validation.

The same two-stage fitting procedure was applied to PPv, using its
\SI{300}{\kelvin} compression curve followed by a fit to 32
pressure--temperature--volume points spanning
\SIrange{100}{145}{\giga\pascal} and \SIrange{300}{4000}{\kelvin}. Because
the PPv compression data begin at \SI{100}{\giga\pascal}, its $V_0$,
$K_{T0}$, and $K'_{T0}$ values are extrapolated reference parameters of the
Birch--Murnaghan form and should not be used for zero-pressure prediction.
The fitted parameters are listed in Table~\ref{tab:s4_eos}. Independent
MACE-QHA and DFT-QHA pressure--volume relations at \SI{300}{\kelvin} and
\SI{2000}{\kelvin} are compared in Text~S5 and
Figs.~\ref{fig:s4_mgo_eos}--\ref{fig:s5_ppv_eos}, together with the Pv
comparison in Fig.~2 of the main text.

\begin{table}[!htbp]
\centering
\small
\caption{Quasi-harmonic thermal equation-of-state parameters for MgO,
bridgmanite, and post-perovskite. The MgO and bridgmanite reference-state
parameters were obtained from third-order Birch--Murnaghan fits to the 298~K
compression curves; the PPv fit used the 300~K compression curve over
100--145~GPa. The thermal parameters were obtained from finite-strain Debye
fits with $V_0$, $K_{T0}$, and $K'_{T0}$ held fixed. Uncertainties are formal
$1\sigma$ fitting uncertainties. The extrapolated PPv reference parameters
should be used only for interpolation within its fitted pressure--temperature
range.}
\label{tab:s4_eos}
\begin{tabular}{lcccc}
\toprule
Parameter & \ce{MgO} & Bridgmanite & Post-perovskite & Unit \\
\midrule
$V_0$ &
$11.747\pm0.002$ &
$25.696\pm0.007$ &
$25.24\pm0.10$ &
cm$^3$ mol$^{-1}$ \\
$K_{T0}$ &
$143.03\pm0.25$ &
$218.05\pm0.79$ &
$226.4\pm7.4$ &
GPa \\
$K'_{T0}$ &
$4.181\pm0.005$ &
$4.021\pm0.017$ &
$3.90\pm0.07$ &
-- \\
$\gamma_0$ &
$1.625\pm0.002$ &
$1.687\pm0.007$ &
$1.699\pm0.005$ &
-- \\
$q_0$ &
$1.289\pm0.007$ &
$1.942\pm0.026$ &
$1.842\pm0.019$ &
-- \\
$\theta_{D0}$ &
$712.6\pm4.4$ &
$808\pm21$ &
$856\pm4$ &
K \\
Equivalent $V_0$ &
$19.506\pm0.003$ &
$42.670\pm0.013$ &
$41.91\pm0.17$ &
\AA$^3$ f.u.$^{-1}$ \\
\bottomrule
\end{tabular}
\end{table}

\section*{Text S3. LBTE formalism and numerical convergence}

For a sufficiently small temperature gradient, the nonequilibrium occupation
of phonon mode $\lambda$ may be written as
\begin{equation}
n_{\lambda}
=
n_{\lambda}^{0}
-
\frac{\partial n_{\lambda}^{0}}{\partial T}
\mathbf F_{\lambda}\cdot\nabla T,
\label{eq:s3_distribution}
\end{equation}
where $n_{\lambda}^{0}$ is the equilibrium Bose--Einstein distribution and
$\mathbf F_{\lambda}$ describes the linear response of mode $\lambda$ to the
applied temperature gradient. The corresponding steady-state linearized
phonon Boltzmann equation is
\begin{equation}
\mathbf v_{\lambda}\cdot\nabla T
\frac{\partial n_{\lambda}^{0}}{\partial T}
=
\left(
\frac{\partial n_{\lambda}}{\partial t}
\right)_{\mathrm{coll}},
\label{eq:s3_lbte}
\end{equation}
where the right-hand side is the phonon collision operator. It contains
three-phonon scattering and, in the isotope calculations for \ce{MgO}, the
additional phonon--isotope scattering contribution. The
thermal-conductivity tensor is defined in Eq.~(2) of the main text. Within
the single-mode relaxation-time approximation,
\begin{equation}
F_{\lambda,\beta}
=
\tau_{\lambda}v_{\lambda,\beta},
\label{eq:s3_rta}
\end{equation}
where $\tau_{\lambda}$ is the total mode relaxation time. In the direct
solution, $\mathbf F_{\lambda}$ is instead obtained by solving the complete
linearized collision system, thereby retaining coupling between phonon modes
\citep{Chaput2013,Togo2015}. The direct-LBTE and RTA conductivities were
obtained from the same force constants and numerical settings; the RTA values
were used only for comparison.

\subsection*{Numerical convergence}

Third-order force constants were constructed with Phono3py~4.4.0 without a
pair-distance cutoff \citep{Togo2015,Togo2023}. The production supercells
contained 250 atoms and 842 displaced structures for \ce{MgO}, 160 atoms and
14\,675 displaced structures for Pv, and 320 atoms and 15\,440 displaced
structures for PPv.

Supercell convergence was evaluated at representative conditions for every
phase (Table~\ref{tab:s5_supercells}). For \ce{MgO}, supercells from
$2\times2\times2$ to $6\times6\times6$ were tested, and the remaining
change between $5\times5\times5$ and $6\times6\times6$ was below 3\% at both
test conditions. For Pv, enlarging the supercell from $2\times2\times2$ to
$3\times3\times3$ changed the individual diagonal components by at most
0.7\%. For PPv, the $4\times2\times2$, $5\times2\times2$, and
$6\times2\times2$ cells contained 320, 400, and 480 atoms and generated
15\,440, 19\,104, and 22\,768 third-order displaced structures,
respectively. At \SI{135}{\giga\pascal} and \SI{4000}{\kelvin}, using a
fixed $10\times10\times10$ transport mesh, the corresponding
orientationally averaged conductivities were 5.85, 5.83, and
\SI{5.84}{\watt\per\metre\per\kelvin}. The production
$4\times2\times2$ value differed from the $6\times2\times2$ result by
0.12\%, and all three tested cells agreed within 0.22\%.

\begin{table}[!htbp]
\centering
\small
\caption{Summary of the supercell convergence tests. The reported change is
relative to the larger comparison cell.}
\label{tab:s5_supercells}
\begin{tabular}{lccccl}
\toprule
Phase & $P$ (GPa) & $T$ (K) & Production cell & Comparison cell & Change \\
\midrule
\ce{MgO} & 0 & 300 & $5\times5\times5$ & $6\times6\times6$ & 2.8\% \\
\ce{MgO} & 135 & 3500 & $5\times5\times5$ & $6\times6\times6$ & 2.2\% \\
Pv & 120 & 3200 & $2\times2\times2$ & $3\times3\times3$ & 0.2\% \\
PPv & 135 & 4000 & $4\times2\times2$ & $6\times2\times2$ & 0.1\% \\
\bottomrule
\end{tabular}
\end{table}

\begin{figure}[!htbp]
  \centering
  \includegraphics[width=0.88\textwidth]
  {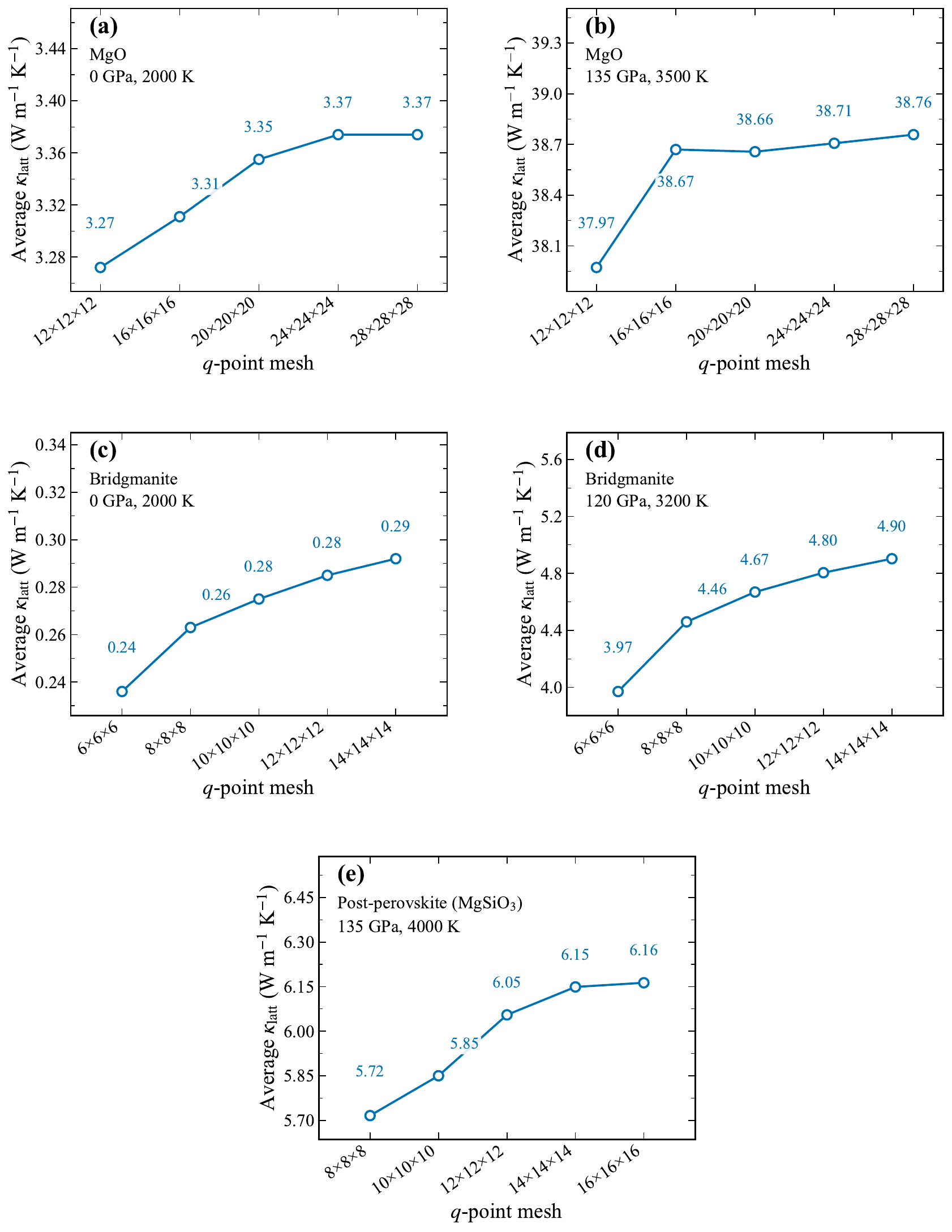}
  \caption{Reciprocal-space convergence of the direct-LBTE lattice thermal
  conductivity of (a) \ce{MgO} at \SI{0}{\giga\pascal} and
  \SI{2000}{\kelvin}, (b) \ce{MgO} at \SI{135}{\giga\pascal} and
  \SI{3500}{\kelvin}, (c) bridgmanite at \SI{0}{\giga\pascal} and
  \SI{2000}{\kelvin}, (d) bridgmanite at \SI{120}{\giga\pascal} and
  \SI{3200}{\kelvin}, and (e) post-perovskite at
  \SI{135}{\giga\pascal} and \SI{4000}{\kelvin}.}
  \label{fig:s3_mesh}
\end{figure}

For reciprocal-space convergence, production meshes of
$24\times24\times24$, $12\times12\times12$, and
$14\times14\times14$ were selected for \ce{MgO}, Pv, and PPv, respectively.
The full mesh comparisons are shown in Fig.~\ref{fig:s3_mesh}, and the PPv
values are listed in Table~\ref{tab:s6_ppv_mesh}. Refining the \ce{MgO} mesh
from $24^3$ to $28^3$ changes the orientationally averaged conductivity by
less than 0.1\% at \SI{0}{\giga\pascal} and \SI{2000}{\kelvin}, and by
0.1\% at \SI{135}{\giga\pascal} and \SI{3500}{\kelvin}. For Pv, increasing
the mesh from $12^3$ to $14^3$ changes the orientational average by 2.4\% at
\SI{0}{\giga\pascal} and \SI{2000}{\kelvin}, and by 2.0\% at
\SI{120}{\giga\pascal} and \SI{3200}{\kelvin}.

For PPv, the orientational average obtained with the selected
$14\times14\times14$ mesh differs from the $16\times16\times16$ result by
0.2\%. The corresponding changes in the [100], [010], and [001] components
are 0.4\%, 0.7\%, and 0.6\%, respectively. The selected mesh therefore
converges every diagonal PPv component to within 0.7\%.

\begin{table}[!htbp]
\centering
\small
\caption{PPv transport-mesh convergence at \SI{135}{\giga\pascal} and
\SI{4000}{\kelvin}, using the $4\times2\times2$ third-order-force-constant
supercell containing 320 atoms. Conductivities are in
\si{\watt\per\metre\per\kelvin}; $\kappa_{\mathrm{latt,avg}}$ is the
arithmetic mean of the three diagonal components.}
\label{tab:s6_ppv_mesh}
\begin{tabular}{ccccc}
\toprule
$q$-point mesh & $\kappa_{[100]}$ & $\kappa_{[010]}$ &
$\kappa_{[001]}$ & $\kappa_{\mathrm{latt,avg}}$ \\
\midrule
$8\times8\times8$   & 7.34 & 4.16 & 5.65 & 5.72 \\
$10\times10\times10$ & 7.64 & 4.16 & 5.75 & 5.85 \\
$12\times12\times12$ & 7.99 & 4.29 & 5.88 & 6.05 \\
$14\times14\times14$ & 8.12 & 4.29 & 6.04 & 6.15 \\
$16\times16\times16$ & 8.15 & 4.26 & 6.07 & 6.16 \\
\bottomrule
\end{tabular}
\end{table}

\section*{Text S4. Conductivity parameterizations and geotherm construction}

The complete pressure--temperature grids used in the direct-LBTE calculations
are listed in Table~\ref{tab:s8_states}. The \ce{MgO} calculations span
\SIrange{0}{135}{\giga\pascal} and \SIrange{300}{3800}{\kelvin}, with the
maximum temperature varying with pressure. The Pv calculations span
\SIrange{0}{135}{\giga\pascal} and \SIrange{300}{3500}{\kelvin}. The PPv
grid contains 32 states spanning \SIrange{100}{145}{\giga\pascal} and
\SIrange{300}{4000}{\kelvin}.

The density--temperature relation defined in the main text was fitted by
least squares in logarithmic space. Only states at \SI{1500}{\kelvin} and
above were included, at 24--\SI{135}{\giga\pascal} for \ce{MgO} and Pv and
100--\SI{145}{\giga\pascal} for PPv. The fits were used only for
interpolation and geotherm evaluation; the direct-LBTE values remain the
primary data. Table~\ref{tab:s7_fit} reports the fitted coefficients and
diagnostics.

\begin{table}[!htbp]
\centering
\small
\caption{Density--temperature parameterizations of \kappalatt. Parameter
uncertainties are $1\sigma$ regression standard errors.}
\label{tab:s7_fit}
\begin{tabular}{lccc}
\toprule
Parameter/statistic & \ce{MgO} & Bridgmanite & Post-perovskite \\
\midrule
Pressure range (GPa) & 24--135 & 24--135 & 100--145 \\
Temperature range (K) & 1500--3800 & 1500--3500 & 1500--4000 \\
Number of fitted points, $N$ & 32 & 31 & 27 \\
$\rho_0$ (g cm$^{-3}$) & 4.5 & 5.0 & 5.0 \\
$T_0$ (K) & 2000 & 2000 & 2000 \\
$\kappa_{\mathrm{latt},0}$ (W m$^{-1}$ K$^{-1}$) &
$43.3\pm0.7$ & $6.87\pm0.10$ & $9.64\pm0.02$ \\
Density exponent, $a$ &
$5.20\pm0.13$ & $5.61\pm0.14$ & $4.99\pm0.04$ \\
Temperature exponent, $b$ &
$1.01\pm0.05$ & $0.97\pm0.04$ & $1.01\pm0.00$ \\
$R^2$ (linear space) & 0.98 & 0.98 & 0.99 \\
RMSE (W m$^{-1}$ K$^{-1}$) & 2.63 & 0.28 & 0.06 \\
MAPE (\%) & 6.0 & 4.8 & 0.4 \\
Maximum absolute residual (\%) & 23.4 & 19.2 & 1.1 \\
\bottomrule
\end{tabular}
\end{table}

The largest parameterization residuals occur at the low-pressure edge of the
fitted domains: at \SI{24}{\giga\pascal} and \SI{3000}{\kelvin} for
\ce{MgO}, and at \SI{24}{\giga\pascal} and \SI{2500}{\kelvin} for Pv. The
direct-LBTE calculations remain the primary results; the parameterizations
provide compact interpolations of the calculated datasets.

\begin{longtable}{
>{\raggedright\arraybackslash}p{0.18\textwidth}
>{\centering\arraybackslash}p{0.13\textwidth}
>{\raggedright\arraybackslash}p{0.58\textwidth}}
\caption{Pressure--temperature states used in the direct-LBTE conductivity
calculations.}
\label{tab:s8_states}\\
\toprule
Phase & $P$ (GPa) & Temperatures (K) \\
\midrule
\endfirsthead
\toprule
Phase & $P$ (GPa) & Temperatures (K) \\
\midrule
\endhead

\ce{MgO} & 0 &
300, 500, 1000, 1500, 2000 \\
& 24 &
300, 500, 1000, 1500, 1900, 2000, 2500, 3000 \\
& 50 &
300, 500, 1000, 1500, 2000, 2200, 2500, 3000 \\
& 75 &
300, 500, 1000, 1500, 2000, 2500, 3000, 3400 \\
& 100 &
300, 500, 1000, 1500, 2000, 2500, 3000, 3500 \\
& 120 &
300, 500, 1000, 1500, 1900, 2000, 2500, 3000, 3500 \\
& 135 &
300, 500, 1000, 1500, 2000, 2500, 3000, 3500, 3800 \\
\midrule
Pv & 0 &
300, 500, 1000, 1500, 2000, 2500, 3000, 3500 \\
& 24 &
300, 500, 1000, 1500, 1900, 2500, 3000, 3500 \\
& 50 &
300, 500, 1000, 1500, 2000, 2200, 2500, 3000, 3500 \\
& 75 &
300, 500, 1000, 1500, 2000, 2600, 3000, 3500 \\
& 100 &
300, 500, 1000, 1500, 2000, 2500, 3000, 3500 \\
& 120 &
300, 500, 1000, 1500, 2000, 2500, 3000, 3200, 3500 \\
& 135 &
300, 500, 1000, 1500, 2000, 2500, 3000, 3200, 3500 \\
\midrule
PPv & 100 &
300, 1500, 2000, 2500, 3000, 3500 \\
& 110 &
300, 1500, 2000, 2500, 3000, 3500 \\
& 120 &
300, 1500, 2000, 2500, 3000, 3500 \\
& 135 &
300, 1500, 2000, 2500, 3000, 3500, 4000 \\
& 145 &
300, 1500, 2000, 2500, 3000, 3500, 4000 \\
\bottomrule
\end{longtable}

The lower-mantle geotherm of \citet{StaceyDavis2008} was represented using
shape-preserving piecewise-cubic Hermite interpolants for pressure and
temperature as functions of depth and for depth as a function of pressure.
The tabulated entry at a depth of \SI{1571}{\kilo\metre}
($P=\SI{62.5}{\giga\pascal}$) is inconsistent with the adjacent pressure
values and was excluded before interpolation. The retained data span
\SIrange{24}{135.7}{\giga\pascal} and
\SIrange{1933}{3739}{\kelvin}.

At each point along the resulting profile, the density of each phase was
calculated from its thermal equation of state
(Table~\ref{tab:s4_eos}) and inserted into the corresponding
density--temperature conductivity parameterization
(Table~\ref{tab:s7_fit}). The 95\% confidence intervals include the parameter
covariances of both the conductivity and thermal-EOS fits. The cold and
thermal EOS covariance blocks were propagated separately because their
cross-covariance was not estimated in the two-stage fitting procedure. The
intervals do not include uncertainties in the adopted geotherm, the phase
boundary, or the underlying transport model.

\section*{Text S5. Additional validation of the fine-tuned potentials}

\begin{figure}[!htbp]
  \centering
  \includegraphics[width=0.78\textwidth]{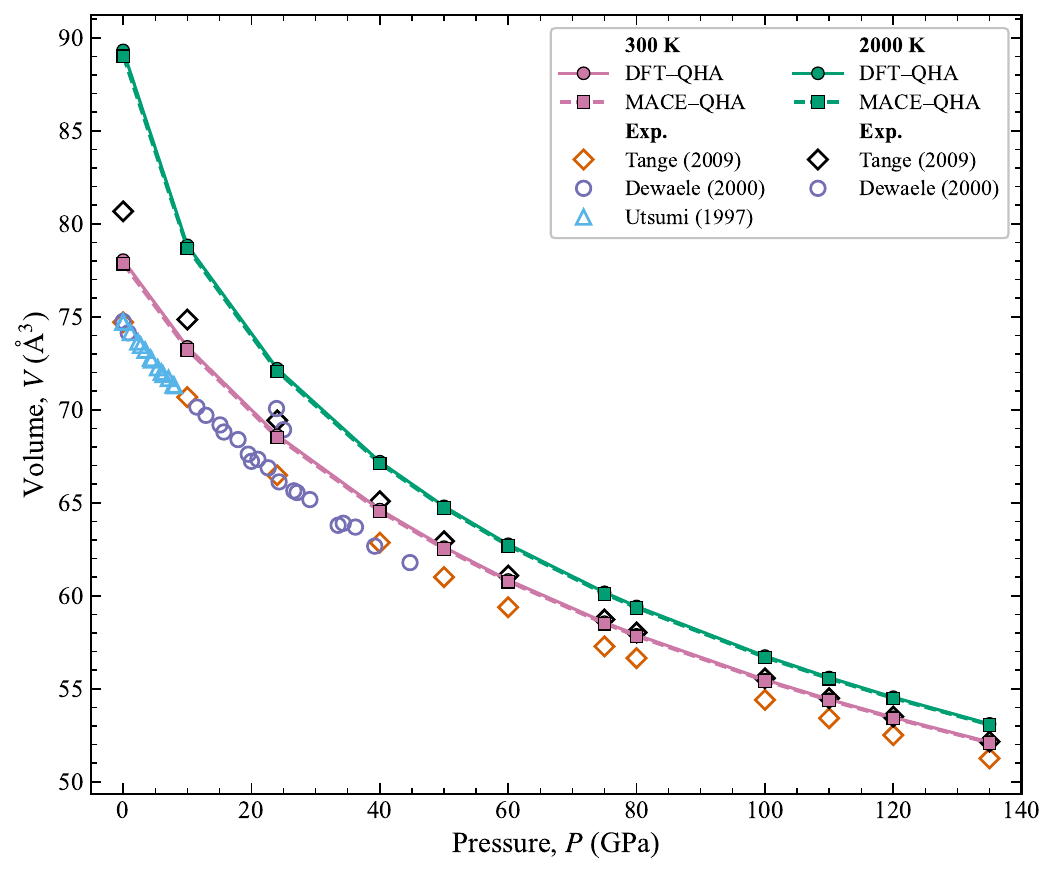}
  \caption{Quasi-harmonic pressure--volume relations for MgO at 300 and
  2000~K. MACE-QHA and DFT-QHA results are compared with the measurements
  and reference equation of state of \citet{Dewaele2000}, \citet{Utsumi1998},
  and \citet{Tange2009}. Volumes are given per conventional cell containing
  four formula units ($Z=4$).}
  \label{fig:s4_mgo_eos}
\end{figure}

\begin{figure}[!htbp]
  \centering
  \includegraphics[width=0.78\textwidth]{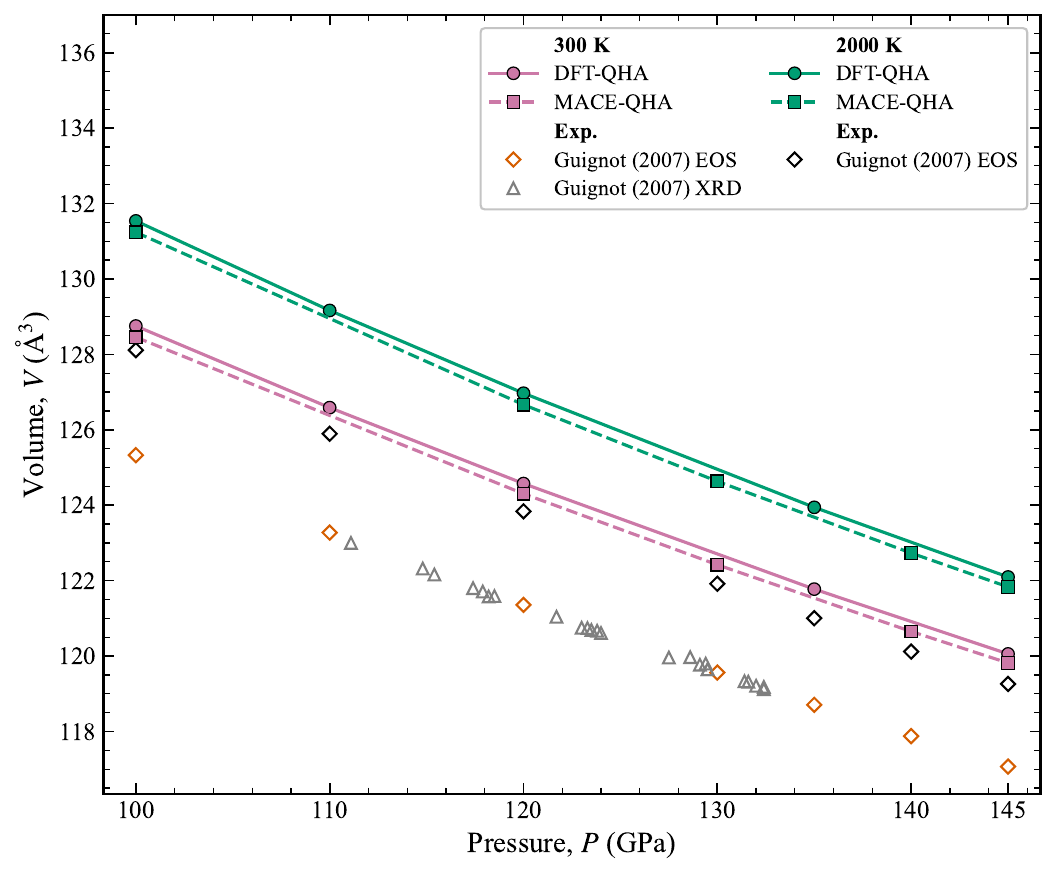}
  \caption{Quasi-harmonic pressure--volume relations for post-perovskite at
  300 and 2000~K. MACE-QHA and DFT-QHA results are compared with the thermal
  equation of state of \citet{Guignot2007} evaluated at 300 and 2000~K and
  with their 300~K diffraction data. Their high-temperature diffraction data
  and those of \citet{Komabayashi2008} were collected at varying temperatures
  and are not plotted. Volumes are given per conventional cell containing
  four formula units ($Z=4$).}
  \label{fig:s5_ppv_eos}
\end{figure}

The MACE-QHA pressure--volume relations for MgO and PPv are compared with the
corresponding DFT-QHA results at 300 and \SI{2000}{\kelvin} in
Figs.~\ref{fig:s4_mgo_eos} and~\ref{fig:s5_ppv_eos}. These comparisons
complement the Pv result shown in the main text. Across the tested
pressure--temperature ranges, the maximum MACE--DFT volume differences are
approximately 0.35\% for MgO and 0.24\% for PPv.

For \ce{MgO}, room-temperature compression measurements and a published
high-pressure thermal equation of state provide external references
\citep{Dewaele2000,Utsumi1998,Tange2009}. For PPv, the thermal equation of
state of \citet{Guignot2007}, evaluated at \SI{300}{\kelvin} and
\SI{2000}{\kelvin}, together with their \SI{300}{\kelvin} diffraction data,
provides the experimental reference.

\begin{figure}[!p]
  \centering
  \includegraphics[width=0.60\textwidth]
  {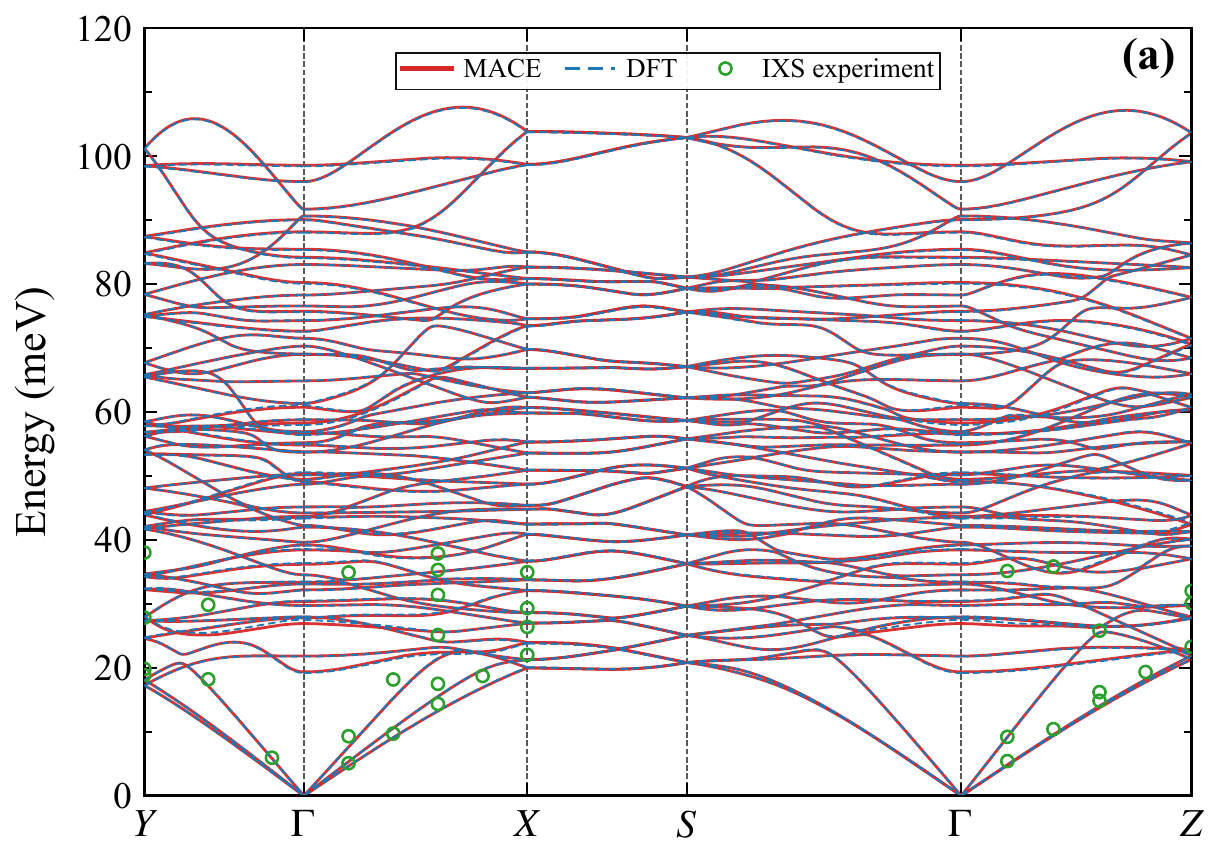}
  \par\vspace{6pt}

  \includegraphics[width=0.60\textwidth]
  {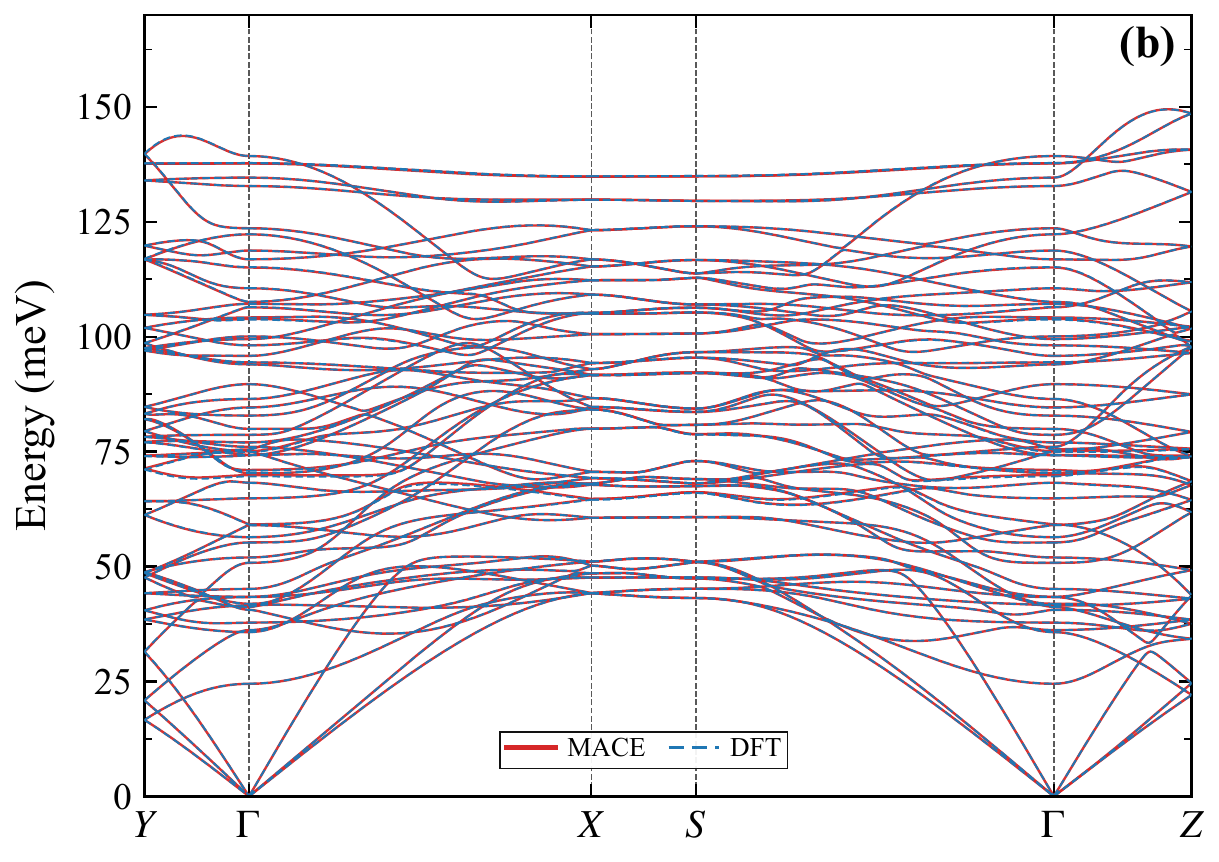}
  \par\vspace{6pt}

  \includegraphics[width=0.60\textwidth]
  {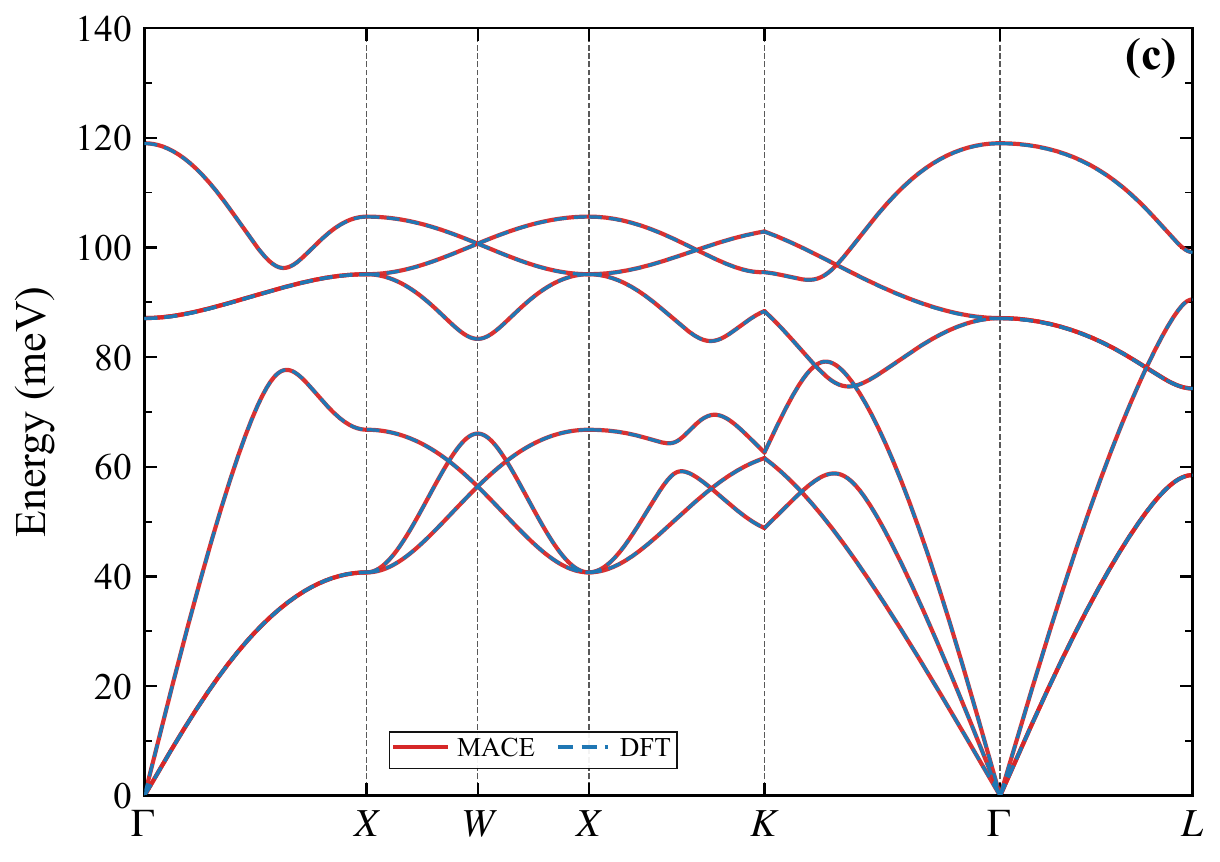}

  \caption{Phonon dispersions obtained from MACE and DFT force constants for
  (a) bridgmanite at \SI{0}{\giga\pascal} and \SI{300}{\kelvin};
  (b) post-perovskite at \SI{135}{\giga\pascal} and
  \SI{3500}{\kelvin}; and (c) \ce{MgO} at
  \SI{120}{\giga\pascal} and \SI{3200}{\kelvin}. In panel (a), the
  calculations are compared with the ambient-condition inelastic
  X-ray-scattering measurements of \citet{Wehinger2016}. All calculations
  were performed at the corresponding QHA equilibrium structures. Solid and
  dashed curves denote the MACE and DFT results, respectively. For the
  compressed-state comparisons in panels (b) and (c), the RMS frequency
  differences between MACE and DFT are 0.014 and
  \SI{0.008}{\tera\hertz}, respectively. Frequencies are plotted in meV
  (\SI{1}{\tera\hertz}=\SI{4.136}{\milli\electronvolt}).}
  \label{fig:s6_phonons}
\end{figure}

Phonon dispersions provide an additional validation target that was not
fitted directly. The main text shows the compressed-state Pv comparison at
\SI{100}{\giga\pascal} and \SI{3000}{\kelvin}, together with the
ambient-condition \ce{MgO} comparison. Figure~\ref{fig:s6_phonons} shows
the ambient-condition Pv comparison, the compressed-state PPv comparison at
\SI{135}{\giga\pascal} and \SI{3500}{\kelvin}, and the compressed-state
\ce{MgO} comparison at \SI{120}{\giga\pascal} and
\SI{3200}{\kelvin}. The RMS frequency differences at the three compressed
states are 0.008, 0.016, and \SI{0.014}{\tera\hertz} for \ce{MgO}, Pv, and
PPv, respectively, and neither the MACE nor DFT dispersions contain unstable
branches. Frequencies were sorted at each sampled wave vector before
numerical differencing to avoid arbitrary branch permutations at crossings.

\clearpage

\section*{Text S6. Directional conductivity of bridgmanite and
post-perovskite}

Both \ce{MgSiO3} polymorphs exhibit anisotropic lattice thermal conductivity.
For comparison across the pressure series, the three diagonal components are
reported in rank order as $\kappa_{\min}$, $\kappa_{\mathrm{mid}}$, and
$\kappa_{\max}$. This convention does not affect the orientational average
or the anisotropy ratio $\kappa_{\max}/\kappa_{\min}$. For PPv in the
\textit{Cmcm} setting, the rank order corresponds to [010], [001], and
[100], respectively; [010] is normal to the \ce{SiO6} layers.

For Pv, $\kappa_{\max}/\kappa_{\min}$ at \SI{300}{\kelvin} increases from
1.09 at \SI{24}{\giga\pascal} to 1.23, 1.29, and 1.33 at 100, 120, and
\SI{135}{\giga\pascal}, respectively. At \SI{135}{\giga\pascal} and
\SI{3500}{\kelvin}, the ratio is 1.23. The normalized spread,
$(\kappa_{\max}-\kappa_{\min})/\kappalatt$, is 20\% at
\SI{100}{\giga\pascal} and 28\% at \SI{135}{\giga\pascal}, comparable to
the approximately 25\% anisotropy reported by \citet{Ghaderi2017}.

PPv is more strongly anisotropic. At \SI{300}{\kelvin},
$\kappa_{[100]}>\kappa_{[001]}>\kappa_{[010]}$ throughout the investigated
pressure range, consistent with \citet{DekuraTsuchiya2019}. The ratio
$\kappa_{[100]}/\kappa_{[010]}$ decreases modestly from 1.76 at
\SI{100}{\giga\pascal} to 1.70 at \SI{145}{\giga\pascal}. Between
1500 and \SI{4000}{\kelvin}, it ranges from 1.82 to 2.06.

At \SI{135}{\giga\pascal} and \SI{3500}{\kelvin}, the [100], [010], and
[001] components are 1.32, 0.70, and 0.98 times the orientational average,
respectively. Relative to the polycrystalline average used by
\citet{DekuraTsuchiya2019}, defined as the mean of the arithmetic and
harmonic averages, the corresponding values are 1.36, 0.72, and 1.01,
close to their reported values of approximately 1.4, 0.7, and 0.98. The
ratio $\kappa_{[100]}/\kappa_{[001]}$ is 1.34--1.40, consistent with the
approximately 40\% difference between these directions reported by
\citet{Ammann2014}. PPv therefore remains substantially more anisotropic
than Pv throughout the investigated pressure--temperature range.

\section*{Text S7. Bridgmanite--post-perovskite phase stability}

The Pv--PPv phase boundary was determined from separate DFT-QHA Gibbs free
energies for the two polymorphs, normalized per \ce{MgSiO3} formula unit.
Both phases were calculated using identical PAW datasets and a common
plane-wave cutoff of \SI{650}{\electronvolt}. Brillouin-zone sampling was
converged separately for each polymorph and held fixed across its pressure
and volume series. Harmonic vibrational free energies were calculated using
$2\times2\times2$ supercells of the 20-atom Pv and PPv cells, containing
160 atoms, with a displacement amplitude of \SI{0.01}{\angstromunit} and a
$30\times30\times30$ phonon $q$-point mesh.

The Gibbs free-energy difference was defined as
\begin{equation}
\Delta G(P,T)
=
G_{\mathrm{PPv}}(P,T)-G_{\mathrm{Pv}}(P,T),
\label{eq:s7_delta_g}
\end{equation}
such that $\Delta G<0$ denotes PPv stability. Gibbs free energies were
evaluated at pressures of 110, 112, 115, 118, 120, 123, 125, and
\SI{130}{\giga\pascal} between 0 and \SI{3900}{\kelvin}, in
\SI{100}{\kelvin} increments. At each temperature, the transition pressure
was obtained by interpolating the zero crossing of $\Delta G$ with pressure.
A linear boundary was fitted to the transition pressures between
\SI{2000}{\kelvin} and \SI{3500}{\kelvin}.

Over this temperature interval, the transition pressure increases from
\SI{111.7}{\giga\pascal} to \SI{122.4}{\giga\pascal} and is represented by
\begin{equation}
P_{\mathrm{tr}}(T)
=
96.62+0.00736\,T,
\label{eq:s7_boundary}
\end{equation}
where $P_{\mathrm{tr}}$ is in GPa and $T$ is in K. The corresponding
Clapeyron slope is
\SI{7.36 \pm 0.13}{\mega\pascal\per\kelvin}
($1\sigma$; rounded to $7.4\pm0.1$ in the main text). The regression has
$R^2=0.996$ and an RMSE of \SI{0.23}{\giga\pascal}. The adopted geotherm
intersects the fitted boundary at \SI{116.2}{\giga\pascal} and
\SI{2660}{\kelvin}, corresponding to a depth of approximately
\SI{2550}{\kilo\metre}.

\begin{figure}[!htbp]
  \centering
  \includegraphics[width=0.75\textwidth]
  {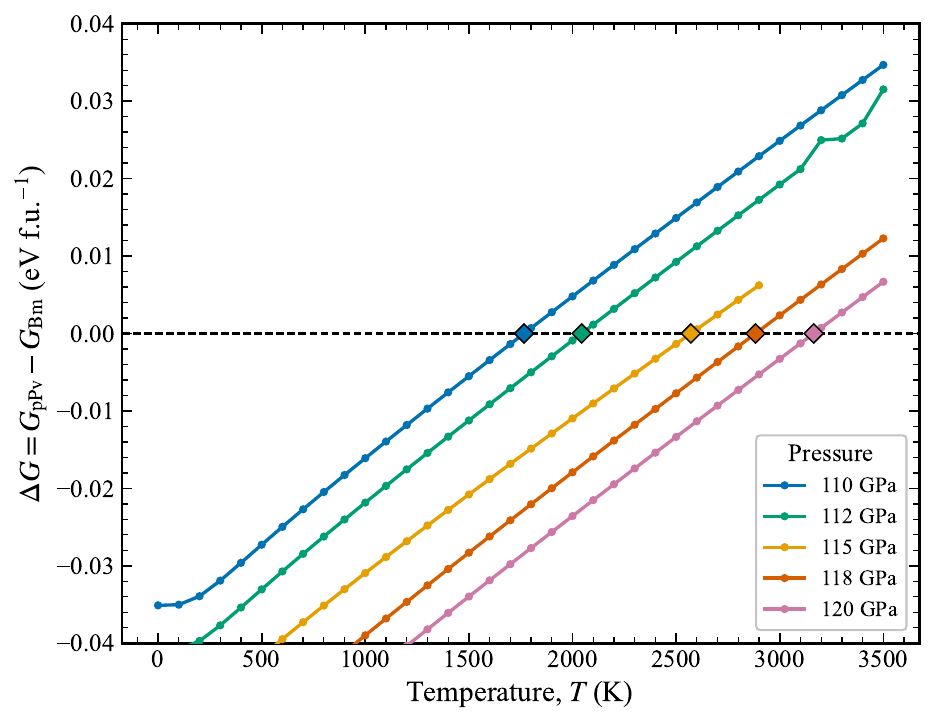}
  \caption{DFT-QHA Gibbs free-energy difference
  $\Delta G=G_{\mathrm{PPv}}-G_{\mathrm{Pv}}$ as a function of temperature at
  selected pressures. The horizontal dashed line marks $\Delta G=0$, and
  diamonds identify the interpolated coexistence temperatures. Negative
  $\Delta G$ denotes post-perovskite stability.}
  \label{fig:s7_boundary}
\end{figure}

The quasi-harmonic boundary does not include intrinsic anharmonic
free-energy contributions that can curve the transition at high temperature
\citep{ZhangWentzcovitch2022}. Along the adopted geotherm, Pv conductivities
calculated at pressures above the transition are therefore treated as
metastable, as indicated by the dashed segments in Fig.~9 of the main text.
The Gibbs free-energy differences and their interpolated zero crossings are
shown in Fig.~\ref{fig:s7_boundary}.

\bibliographystyle{plainnat}
\bibliography{references}